\documentclass[prb,preprint,showpacs,preprintnumbers,amsmath,amssymb]{revtex4}
\usepackage{graphicx}
\usepackage{bm}

\newcommand{\PR}[1]{Phys. Rev. B {\bf {#1}}}
\newcommand{\PRB}[1]{Phys. Rev. B {\bf {#1}}}
\newcommand{\PRL}[1]{Phys.\ Rev.\ Lett. {\bf {#1}}}
\newcommand{\JPSJ}[1]{J.\ Phys.\ Soc.\ Jpn. {\bf #1}}
\newcommand{\sgt}{$\raisebox{-0.6ex}{$\stackrel{>}{\sim}$}$}
\newcommand{\slt}{$\raisebox{-0.6ex}{$\stackrel{<}{\sim}$}$}
\newcommand{\bvec}[1]{\mbox{\boldmath $#1$}}
\newcommand{\D}{\delta }

\newcommand{\vq}{\bvec{q}}

\newcommand{\vk}{\bvec{k}}

\newcommand{\vS}{\bvec{S}}

\newcommand{\eq}[1]{eq.~(\ref{#1})}
\newcommand{\fig}[1]{Fig.~\ref{#1}}
\newcommand{\be}{\begin{equation}}
\newcommand{\ee}{\end{equation}}
\newcommand{\bea}{\begin{eqnarray}}
\newcommand{\no}{\nonumber}
\newcommand{\eea}{\end{eqnarray}}
\newcommand{\bean}{\begin{eqnarray*}}
\newcommand{\eean}{\end{eqnarray*}}
\newcommand{\bfi}{\begin{figure}}
\newcommand{\efi}{\end{figure}}
\newcommand{\bc}{\begin{center}}
\newcommand{\ec}{\end{center}}
\newcommand{\ba}{\begin{array}}
\newcommand{\ea}{\end{array}}

\begin{document}

\title{Magnetic Excitation of $\boldsymbol{t}$--$\boldsymbol{J}$ Model with Quasi-One-Dimensional Fermi Surface
--- Possible Relevance to LSCO Systems} 

\author{Hiroyuki Yamase$^{1}$ and Hiroshi Kohno$^{2}$}

\affiliation{
{$^{1}$}Institute for Solid State Physics, University of Tokyo, 
 5-1-5 Kashiwanoha, Kashiwa, Chiba 277-8581, Japan\\ 
{$^{2}$}Graduate School of Engineering Science, 
Osaka University, Toyonaka, Osaka 560-8531, Japan
}

\date{May 14, 2001}

\begin{abstract}
On the basis of the picture of a quasi-one-dimensional (q-1d) 
Fermi surface (FS), 
recently proposed by authors for LSCO systems,  
spin excitation spectrum, 
${\rm Im} \chi(\vq,\,\omega)$,  is calculated in the \lq RPA'
within the slave-boson mean-field approximation to the $t$--$J$ model. 
It is found that ${\rm Im} \chi(\vq,\,\omega)$ 
shows both incommensurate (IC) and  diagonal IC (DIC) peaks, 
whose realization does not depend on the existence of 
the $d$-wave gap. 
The peak  positions do not change appreciably with 
$\omega$ 
and the sharp peaks survive down to the low hole doping rate. 
The $d$-wave gap suppresses both the IC peak and the DIC peak, 
but the degree of suppression as a function 
of $\omega$ is different between them. 
Taking these results together with results for the two-dimensional FS,  
we argue that essential features of 
magnetic excitation in LSCO systems can be understood 
in terms of the q-1d picture of the FS. 
\end{abstract}


\maketitle

\section{Introduction}
Recently, elastic neutron 
scattering\cite{tranquada1,tranquada2,tranquada3,ichikawa00} and hard 
X-ray scattering\cite{zimmermann,niemoller} for 
La$_{1.6-x}$Nd$_{0.4}$Sr$_{x}$CuO$_4$ (LNSCO) have revealed four 
superlattice peaks at $(0,\,\pm 4\pi\eta)$ and 
$(\pm 4\pi\eta,\,0)$\cite{Dthesis}. These peaks are believed to   
come from some possible charge 
density modulation (CDM), accompanied 
by the magnetic incommensurate peaks at $(\pi,\,\pi\pm 2\pi\eta)$ and 
$(\pi\pm 2\pi\eta,\,\pi)$ at lower temperature. 
Here the \lq incommensurate' means that 
the magnetic peak position is away from the commensurate
position, $(\pi,\,\pi)$, and the value of $\eta$ is called as
incommensurability. 
This experimental data has been often discussed in terms of the so-called 
\lq spin-charge stripes' hypothesis\cite{tranquada1,tranquada2}. 
In this hypothesis, 
one-dimensional (1d) charge order (or its fluctuations), 
namely \lq charge stripes',  
is assumed in each CuO$_{2}$ plane  
and is considered to be essential to the realization of 
the incommensurate magnetic order 
(or its fluctuations); the direction of 
\lq charge stripes' is assumed to alternate 
along the $c$-axis to explain 
the \lq observed' four superlattice peaks.   

On the theoretical side, some studies consider 
the \lq spin-charge stripes' picture as a 
mechanism of superconductivity\cite{emery97,kivelson98}, 
and others use it to explain characteristic features of 
LSCO systems\cite{salkola,tohyama99,machida99}. 
The possible formation of \lq spin-charge stripes', however, has not been 
clarified and has been one of the hot theoretical issues. 

\bfi[th]
\begin{center}
 \includegraphics[width=8cm]{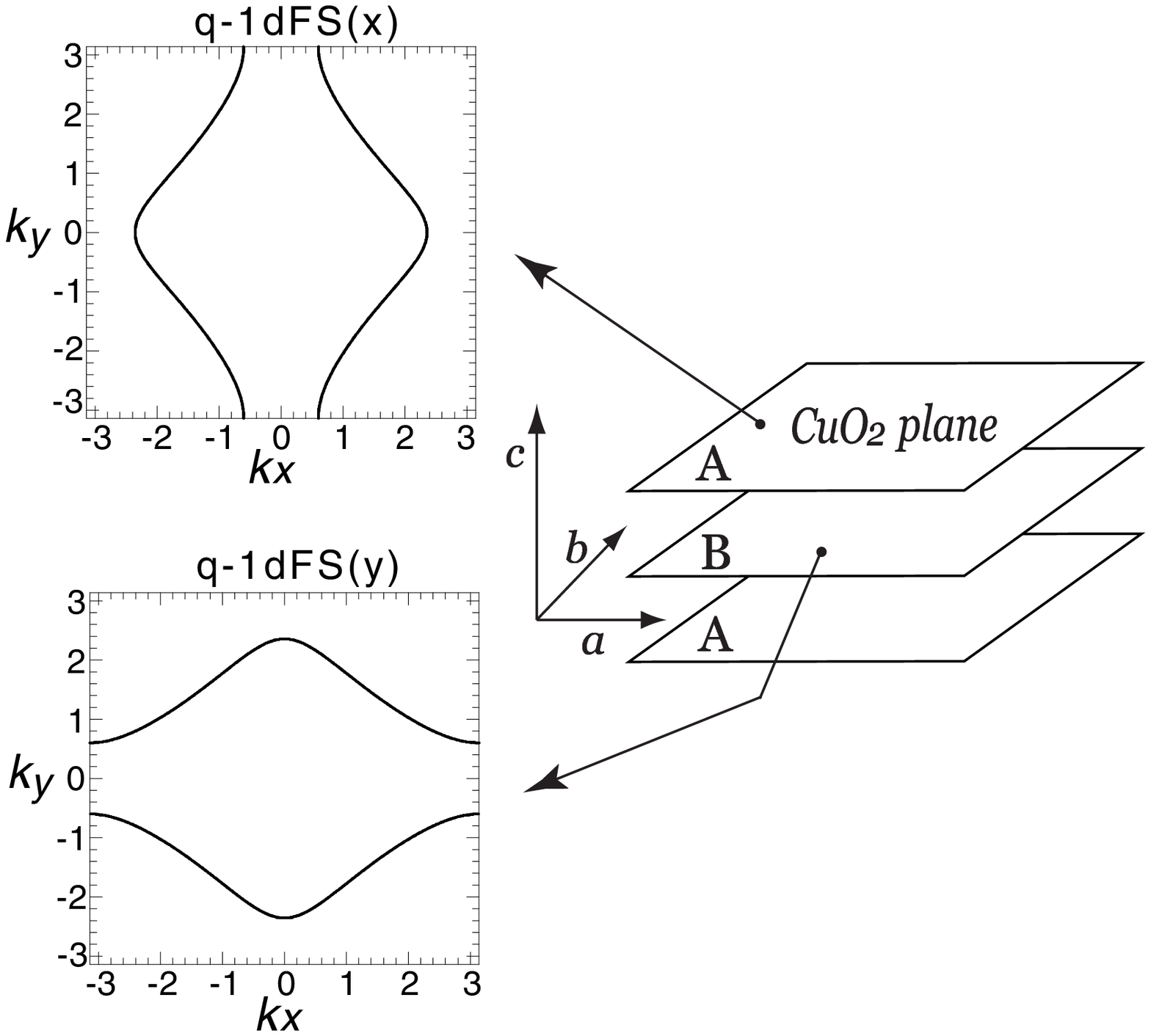}
     \caption{Proposed quasi-one-dimensional picture of the FS. 
     Either of two kinds of FSs, the q-1dFS($x$) or the q-1dFS($y$), 
     is realized in each CuO$_{2}$
     plane and they are stacked alternately along the $c$-axis.} 
    \label{q1dpicture}
\end{center}
\efi

On the other hand, considering that some possible CDM 
has been indicated only in LNSCO with
$x=0.10$, 0.12, 0.15\cite{ichikawa00}, and neither in LNSCO with other
hole density nor in other high-$T_{\rm c}$ cuprates such as 
La$_{2-x}$Sr$_{x}$CuO$_4$ (LSCO) and YBa$_{2}$Cu$_{3}$O$_{6+y}$ (YBCO), 
we have performed theoretical study\cite{yamase1,yamase2,yamase21} 
under the assumption of uniform charge density, leaving a possible
formation of some kind of CDM to a future study. 
As a result,
we have proposed\cite{yamase1} another possible picture for LSCO systems,  
a quasi-one-dimensional (q-1d) picture of 
the Fermi surface (FS). This is illustrated in \fig{q1dpicture}:  
either of  two kinds of the FSs, q-1dFS($x$) or q-1dFS($y$), is 
realized in each CuO$_{2}$ plane 
and they are stacked alternately along the $c$-axis. 
On the basis of this picture, we have argued that 
the apparently contradicting experimental results between 
the angle-resolved photoemission spectroscopy (ARPES)\cite{ino} 
and the inelastic neutron scattering\cite{yamada} will be reconciled. 
As a microscopic support for the q-1d picture for LSCO systems, 
we have shown the followings\cite{yamase2,yamase21}.  
(1) The two-dimensional (2d) (spatial isotropic) $t$--$J$ model 
has an intrinsic instability toward the formation of 
the q-1dFS at low temperature 
and this instability 
is most enhanced for the band parameters appropriate  to 
LSCO systems.  
(2) The q-1d instability is, however,  usually 
masked by the more prominent $d$-wave pairing instability. 
(3) Nonetheless,  the presence of small 
extrinsic spatial anisotropy is 
sufficient for the q-1d state to manifest in the $d$-wave
pairing state; as an  origin of such anisotropy in LSCO systems, we 
may assume the lattice distortion due to the low-temperature
tetragonal (LTT) structure\cite{crawford,buchner,sakita} 
or its fluctuations\cite{thurston89,chlee96,kimura00}. 
(The q-1d instability of the
FS, as well as its competing nature with $d$-wave pairing instability, 
has independently been found in the Hubbard model also by 
Halboth and Metzner\cite{metzner}.)

In this paper, we perform a detailed study of  
magnetic excitation in the framework of 
the q-1d picture of the FS. 
Taking the q-1dFS consistent with ARPES results\cite{ino}, 
we calculate the dynamical spin susceptibility, $\chi(\vq,\, \omega)$, 
especially ${\rm Im} \chi(\vq,\,\omega)$,  in the \lq RPA'    
within the slave-boson mean-field approximation to the $t$--$J$ model. 
We first focus on the calculations for the single CuO$_{2}$ plane and 
neglect the interlayer coupling. 
After describing the formalism in \S2,  
we show in \S3.1 ${\rm Im} \chi(\vq,\,\omega)$ 
at low temperature where 
the $d$-wave singlet resonating-valence-bond ($d$-RVB) state   
is realized. 
At the same temperature 
we also perform the calculations assuming 
the uniform  RVB (u-RVB) state (the state without the $d$-wave gap)    
and study effects of the $d$-wave
gap (\S3.2).  
Effects of the interlayer hopping (the formalism being  given in 
Appendix~A) and thermal fluctuations are 
investigated in \S3.3 and \S3.4, respectively. 
We also study in \S3.5 magnetic excitation for the 2dFS. 
Taking these results, we argue in \S 4 
that essential features of magnetic excitation in LSCO systems can be 
understood in terms of the present q-1d picture of the FS. 
Our argument 
is different from the \lq spin-charge stripes' 
scenario\cite{tranquada1,tranquada2,tranquada3} where 
it is the formation of \lq charge stripes', not effects of the FS,   
that gives rise to the magnetic incommensurate peaks. 

\section{Model and Formalism}
In this section, we give a formalism for a single CuO$_{2}$ plane. 
The case with the interlayer hopping is described in Appendix~A.

\subsection{Mean-field Hamiltonian}
As a theoretical model of high-$T_{\rm c}$ cuprates, we take 
the 2d $t$--$J$ model defined on a square lattice:
\bea
 & &H = -  \sum_{i,\,j,\, \sigma} t\,^{(l)} 
 f_{i\,\sigma}^{\dagger}b_{i}b_{j}^{\dagger}f_{j\,\sigma} + 
   J \sum_{\langle i,j \rangle}  \vS_{i} \cdot \vS_{j}, \label{tJ} \\
& &\hspace{2mm} \sum_{\sigma}f_{i\,\sigma}^{\dagger}f_{i\,\sigma}
          +b_{i}^{\dagger}b_{i}=1  
  {\rm \quad  at\  each\  site}\ i, \label{constraint}
\eea  
where $f_{i\,\sigma}$ ($b_{i}$) is a fermion (boson) operator  
that carries spin $\sigma$ (charge $e$), namely we adopt 
the slave-boson scheme. 
$t^{(l)}$ is the  hopping integral between 
the $l$-th neighbor sites $i$ and $j$ ($l\leq 3$), 
$J >0$ is the superexchange coupling between the nearest neighbor 
spins, and 
$\vS_{i} = \frac{1}{2}\sum_{\alpha,\beta}
   f_{i\,\alpha}^{\dagger}\bvec{\sigma}_{\alpha\,\beta}f_{i\,\beta}$
 with Pauli matrix $\bvec{\sigma}$.   
The constraint \eq{constraint} excludes double occupations at 
every site.  

The q-1dFS can be determined as a 
fully self-consistent solution by introducing 
some small spatial anisotropy in the $t$--$J$ model\cite{yamase2}.  
The resulting q-1dFS, however, is not quantitatively consistent 
with ARPES data 
at each $\D$.  
In this paper, we aim at a {\em semiquantitative} 
study of magnetic excitations
in LSCO systems based on the FS consistent with the ARPES results\cite{ino}.
Therefore, rather than sticking to such self-consistent treatment,  
we take the following phenomenological procedure to reproduce the FS 
consistent with FS segments observed by ARPES at each $\D$. 

We introduce the mean fields: 
$\chi^{(l)}
\equiv \langle \sum_{\sigma}f_{i\,\sigma}^{\dagger}
f_{j \,\sigma}\rangle$, 
$\langle b_{i}^{\dagger}b_{j}\rangle$ and 
$\Delta_{\tau}  
\equiv \langle f_{i\,\uparrow}f_{i+\tau \,\downarrow}- 
f_{i\,\downarrow}f_{i+\tau \,\uparrow}\rangle$, 
where each is taken to be a real constant independent of lattice 
coordinates, and $\tau$ represents 
the nearest neighbor bond direction, namely $\tau=x$ or $y$.  
The local constraint \eq{constraint} is loosened to a global one,   
$\sum_{i}(\sum_{\sigma}f_{i\,\sigma}^{\dagger}f_{i\,\sigma}
          +b_{i}^{\dagger}b_{i})=N$, 
with $N$ being the total number of lattice sites. 
We then decouple the Hamiltonian \eq{tJ} to obtain 
\be
H_{\rm MF} = 
\sum_{\vk, \,\sigma}\xi_{\vk} f_{\vk\,\sigma}^{\dagger} f_{\vk\,\sigma}
+ \sum_{\vk} \Delta_{\vk} \left(f_{-\vk\,\downarrow}^{\dagger} 
f_{\vk\,\uparrow}^{\dagger} + f_{\vk\,\uparrow} 
f_{-\vk\,\downarrow} \right), \label{MFH}
\ee
where 
\bea
\xi_{\vk}&=&\left(F_{x}\cos k_{x}+F_{y} \cos k_{y}\right)+  
F^{'} \cos k_{x} \cos k_{y} +  
F^{''} \left( \cos 2k_{x} +\cos 2k_{y}\right) - \mu  \label{xi} \; ,\\
\Delta_{\vk}&=&-\frac{3}{4}J \Delta_{0}  
\left(\cos k_{x}- \cos k_{y} \right) \; ,\\
F_{x}&=&F_{y}=-2 \left(
      t\!\,^{(1)}\left<b_{i}^{\dagger}b_{j}\right> +
    \frac{3}{8}J \chi^{(1)}\right)\; , \\ 
F^{'}&=&-4 t^{(2)}\left<b_{i}^{\dagger}b_{j}\right> , \quad
F^{''}= -2 t\!\,^{(3)}\left<b_{i}^{\dagger}b_{j}\right>\;,\\
\Delta_{0} &\equiv& \Delta_{x}=-\Delta_{y} \; ,
\eea
and $\mu$ is the chemical potential. We neglect boson degree of freedom,
assuming the condensation to the bottom of its band. 
This assumption will be reasonable at low temperature and 
leads  to $\left<b_{i}^{\dagger}b_{j}\right> \approx \delta$. 

Focusing our attention on LSCO systems, 
we take band parameters as\cite{tanamoto1}  
$t^{(1)}/J =4$, $t^{(2)}/t^{(1)}=-1/6$ and 
$t^{(3)}/t^{(1)}=0$, and determine mean fields self-consistently. 
The resulting \lq 2dFS' turns out to reproduce the observed FS\cite{ino} 
in LSCO with $\delta=0.30$\cite{xD}, but not with lower $\D$ $(\slt 0.22)$.  
In order to reproduce the FS on the basis of the q-1d picture of the FS 
shown in \fig{q1dpicture},  
we reduce for the q-1dFS($x$) the value of $F_{y}$ at each $\D$, 
keeping the other parameters, 
$F_{x}$, $F^{'}$, $F^{''}$, $\mu$ and $\Delta_{0}$, fixed: 
\be
F_{y}=\alpha F_{x} \quad (0 < \alpha \leq 1) \; . \label{alpha}
\ee
The value of $\alpha$ is chosen to adjust our theoretical q-1dFS($x$) 
near $(0,\,\pi)$ to the observed FS segments\cite{ino}. 
The obtained values are 
plotted in \fig{alpha-delta}: the band anisotropy decreases with 
increasing $\delta$ and eventually disappears at $\D=0.30$ where 
the \lq 2dFS' is realized. Note that the \lq 2dFS' was 
used for discussing the LSCO systems in the previous 
theory\cite{tanamoto1,tanamoto2}. 
Hence, the present theory recovers the previous one at 
high $\D$. 

\bfi[ht]
\bc
 \includegraphics[width=7cm]{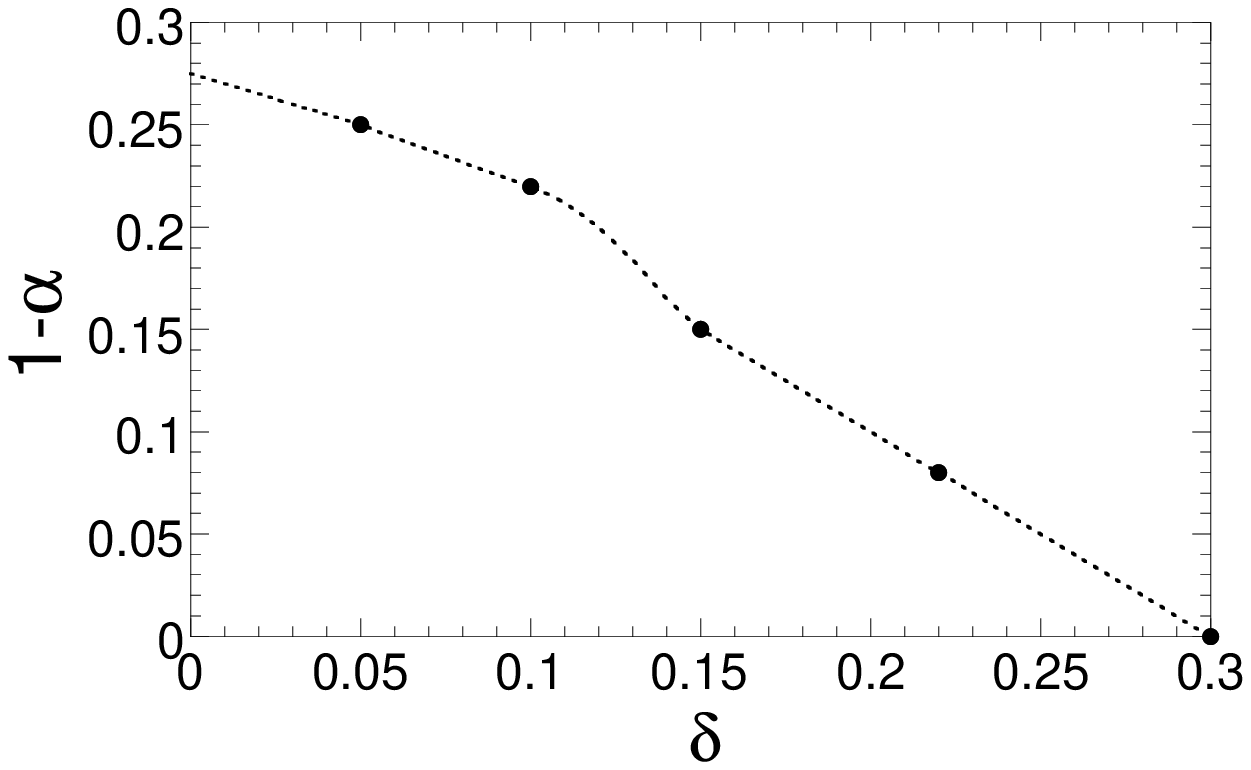}
     \caption{The band anisotropy, $1-\alpha$, determined by adjusting the 
q-1dFS to the observed FS segments for  several choices of 
$\D$. The dotted line 
is drawn smoothly so as to estimate $\alpha$ at each $\D$.}
    \label{alpha-delta}
\ec
\efi

\subsection{Dynamical magnetic susceptibility}
Using the mean-field Hamiltonian, \eq{MFH}, we calculate the 
irreducible dynamical magnetic susceptibility as 
\bea
& &\quad\qquad\chi_{0}(\vq,\,\omega)=\frac{1}{4 N}
\sum_{\vk}\left[C^{+}_{\vk,\,\vk+\vq} 
 \left(\tanh \frac{\beta E_{\vk}}{2}
   -\tanh \frac{\beta E_{\vk +\vq}}{2}\right)
   \frac{1}{E_{\vk}-E_{\vk+\vq}+\omega+{\rm i}\Gamma}\right. \nonumber\\
& &+\left.\frac{1}{2}C^{-}_{\vk,\,\vk+\vq}\left(\tanh \frac{\beta E_{\vk}}{2}
   +\tanh \frac{\beta E_{\vk +\vq}}{2}\right)
\left(\frac{1}{E_{\vk}+E_{\vk+\vq}+\omega+{\rm i}\Gamma}
   +\frac{1}{E_{\vk}+E_{\vk+\vq}-\omega-{\rm i}\Gamma}\right)\right],\,
\eea
where $\beta^{-1}=T$ is temperature and 
\bea
E_{\vk}&=&\sqrt{\xi_{\vk}^{2}+\Delta_{\vk}^{2}} \, , \\
C^{\pm}_{\vk,\,\vk+\vq}&=&\frac{1}{2}
\left(1 \pm \frac{\xi_{\vk}\xi_{\vk+\vq}
 +\Delta_{\vk}\Delta_{\vk+\vq}}{E_{\vk}E_{\vk+\vq}}\right)\; .
\eea
The value of $\Gamma$ should be a positive infinitesimal, but here 
we set $\Gamma=0.01J$, which may simulate finite lifetime of fermions.

The RPA dynamical magnetic susceptibility is then obtained as 
\be
\chi(\vq,\,\omega)=\frac{\chi_{0}(\vq,\,\omega)}{1+2 r
J(\vq)\chi_{0}(\vq,\,\omega)}\; , \label{RPA}
\ee
where $J(\vq)=J (\cos q_{x}+\cos q_{y})$ and we introduce a numerical 
factor $r$ for convenience.  In this RPA, where $r=1$,  
$\chi(\vq,\,0)$ diverges at low temperature in 
the wide doping-region $\delta \slt 0.17$ (in the $d$-RVB state). 
This magnetic instability will be an artifact, 
since such divergence of $\chi(\vq,\,0)$ will be suppressed by 
higher order corrections to $\chi_{0}(\vq,\,\omega)$. 
This aspect we take into account 
phenomenologically by reducing the value of $r$ to 0.35. 
As a result, the divergence of $\chi(\vq,\,0)$ is limited to the region   
$\delta \slt 0.02\; (0.05) $ in the $d$-RVB (u-RVB) state.

\section{Results}
We first present the calculations for a single CuO$_{2}$ plane and 
show the $\vq$-dependence of 
${\rm Im} \chi(\vq,\,\omega)$ for the q-1dFS($x$); temperature is 
set to $T=0.01J$ where the $d$-RVB state is stabilized. 
We also calculate ${\rm Im} \chi(\vq,\,\omega)$ 
assuming the u-RVB state 
at the same temperature 
and study effects of the $d$-wave gap; 
mean-field parameters in the u-RVB 
are determined self-consistently within a manifold of 
$\Delta_{0}\equiv 0$, and  
the same value is taken for $\alpha$ as that in the $d$-RVB state. 
We next investigate effects of the interlayer hopping and  
thermal fluctuations. 
Finally we compare results for the q-1dFS with those for the 2dFS. 
In the following, we take $J$ as an energy unit. 

\subsection{$\vq$-dependence of ${\rm Im} \chi(\vq,\,\omega)$}
\bfi[th]
\bc
 \includegraphics[width=13cm]{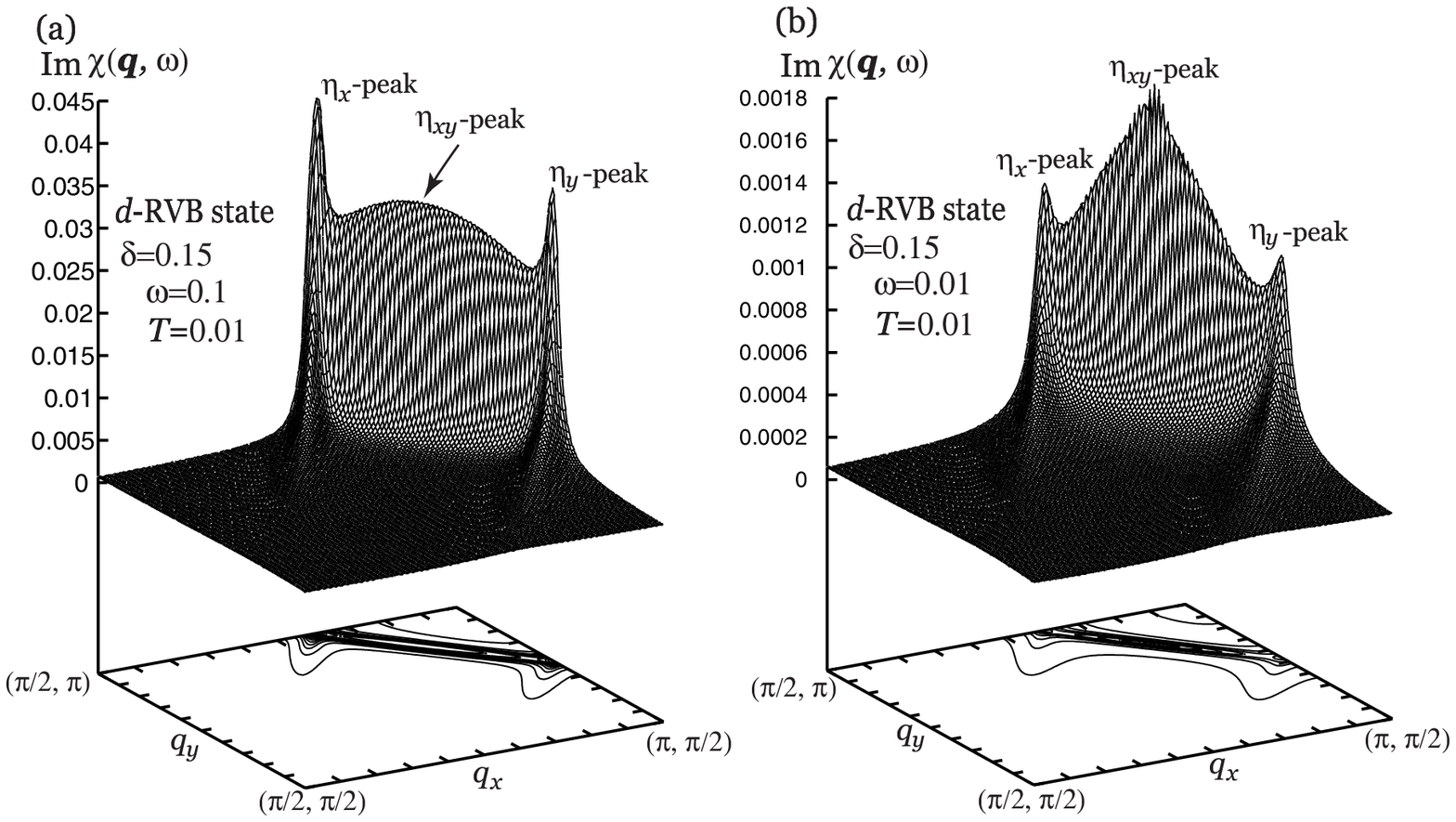}
     \caption{$\vq$-dependence of ${\rm Im} \chi(\vq,\,\omega)$ at
    $\omega=0.1$ (a) and 0.01 (b) in the $d$-RVB state. The contour
    lines     are projected on the $\vq$-plane. 
    In (b), the fine structure 
    around the $\eta_{xy}$-peak is due to the coarse mesh in the plot 
    and should be understood with a smooth interpolation. } 
    \label{3dplot}
\ec
\efi

\bfi[hb]
\bc
 \includegraphics[width=5cm]{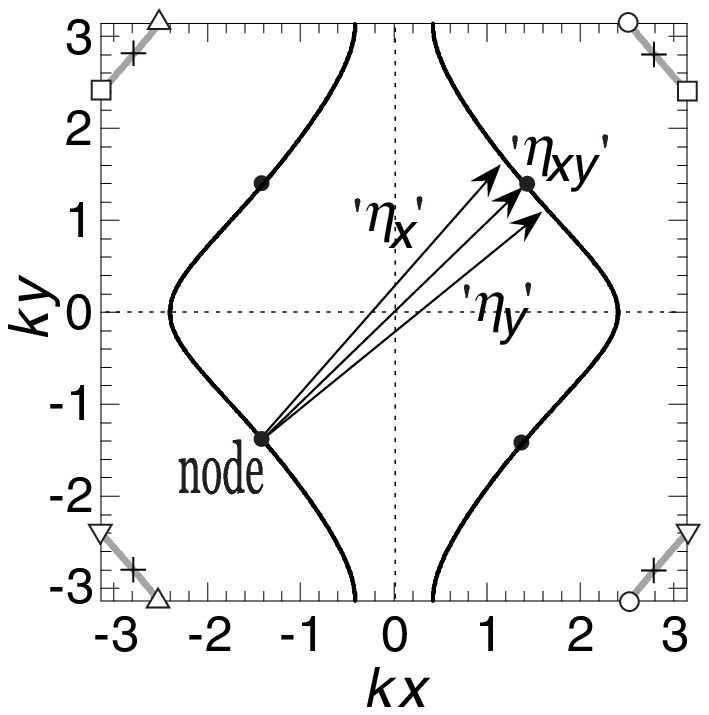}
     \caption{Typical particle-hole scattering processes for the 
q-1dFS($x$) in the $d$-RVB state. 
(The FS is drawn  by setting $\xi_{\vk}=0$, but note that the band
dispersion is given by 
$E_{\vk}=\sqrt{\xi_{\vk}^{2}+\Delta_{\vk}^{2}}$. The
$d$-wave gap nodes on the FS are denoted by the filled circles.)  
The gray line indicates the positions of $\lq 2k_{F}$'-scattering vectors 
connecting the vicinities of the $d$-wave gap nodes. 
The \lq $+$' and the open symbols
denote the location of the $\eta_{xy}$-peak and the
$\eta_{x},\eta_{y}$-peaks, respectively. The same open symbols 
indicate that they are connected by the reciprocal 
lattice vectors. The main scattering vectors 
for the $\eta_{x}$-peak (denoted by \lq $\eta_{x}$'), 
the $\eta_{xy}$-peak (\lq $\eta_{xy}$') 
and the $\eta_{y}$-peak (\lq $\eta_{y}$') are shown explicitly. 
}
    \label{nesting}
\ec
\efi

In \fig{3dplot}(a), we show the $\vq$-dependence of 
${\rm Im} \chi(\vq,\,\omega)$ at $\omega=0.1$ in the 
region $\frac{\pi}{2}\leq q_{x}, q_{y} \leq \pi$ together with the  
projected contour-lines on the $\vq$-plane. 
The overall structure of ${\rm Im} \chi(\vq,\,\omega)$ 
is almost 2d-like even in the 
state with the q-1dFS($x$), except for the absence of  
the exact tetragonal symmetry, 
$(q_{x},\, q_{y}) \rightarrow (\pm q_{y},\, \pm q_{x})$. 
There exist two different incommensurate (IC-) peaks at 
$(\pi-2\pi \eta_{x},\,\pi)$ and $(\pi,\,\pi-2\pi \eta_{y})$, which 
we call the $\eta_{x}$-peak and the $\eta_{y}$-peak, respectively. These 
peaks are connected with each other by a \lq wall', as seen from 
the dense contour lines in \fig{3dplot}(a). 
The center of the \lq wall' located at 
$(\pi-2 \pi \eta_{xy},\,\pi-2 \pi \eta_{xy})$ forms a local 
maximum or a diagonal IC (DIC-) peak, 
which we call the $\eta_{xy}$-peak. 
With decreasing $\omega$, the $\eta_{x},\eta_{y}$-peaks 
are rapidly suppressed compared with the $\eta_{xy}$-peak, 
and the latter then becomes  dominant as shown in \fig{3dplot}(b).

\bfi[ht]
\bc
 \includegraphics[width=8cm]{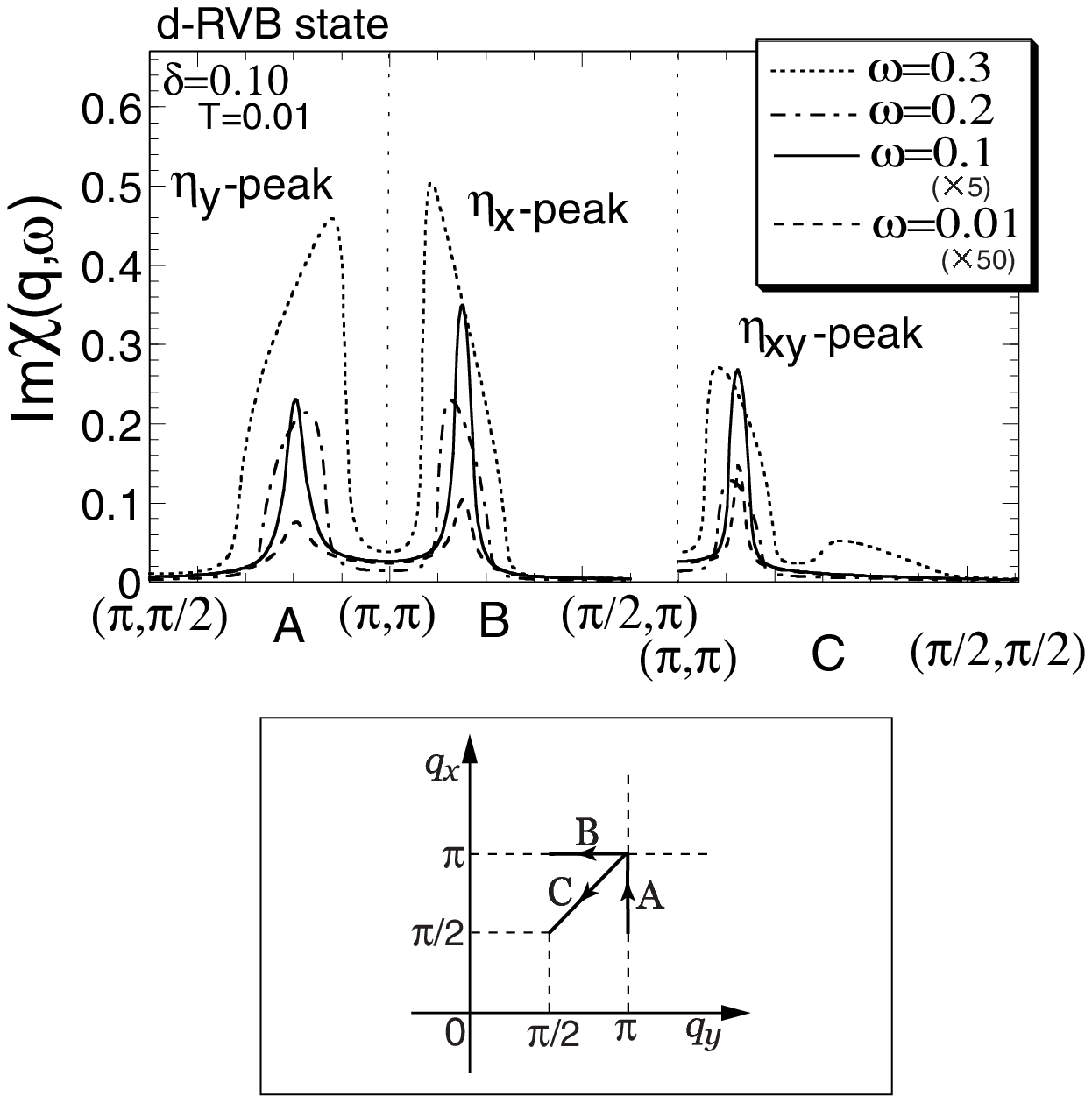}
     \caption{$\vq$-dependence of ${\rm Im} \chi(\vq,\,\omega)$ 
     in the $d$-RVB state at 
     several values of $\omega$ along the
     direction shown in the lower panel. 
     ${\rm Im} \chi(\vq,\,\omega)$ at $\omega=0.1$ and $0.01$ are multiplied 
     by 5 and 50, respectively.}
    \label{wdepend}
\ec
\efi

These structures of ${\rm Im} \chi(\vq,\,\omega)$ 
can be understood in terms of fermiology as follows.   
In the present q-1dFS($x$), $\lq 2k_{F}$'-scattering 
vectors are located on the 
gray lines in \fig{nesting}.
Along this gray line ${\rm Im} \chi(\vq,\,\omega)$ shows  
the \lq wall' structure, and the locations of the $\eta_{xy}$-peak 
are denoted by $\lq +$'   
and those of the $\eta_{x},\eta_{y}$-peaks by the 
open symbols. Since the same open symbols are 
connected by the reciprocal lattice unit, Umklapp processes also 
contribute to the $\eta_{x}, \eta_{y}$-peaks.  
We show in \fig{nesting} the  main scattering processes for each peak around 
$(\pi,\,\pi)$.   
(Umklapp processes are not shown.)  
Each scattering process originates from the $d$-wave gap node.  
In particular, the $\eta_{xy}$-peak results from the scattering 
between the $d$-wave gap nodes. This is why the $\eta_{xy}$-peak becomes 
dominant at lower $\omega$.

To see the $\omega$-dependence of ${\rm Im} \chi(\vq,\,\omega)$ more 
clearly, we perform the $\vq$-scan along three lines, 
each of which is 
across the $\eta_{x}$-peak, the $\eta_{y}$-peak or the
$\eta_{xy}$-peak, respectively, 
and show the result in \fig{wdepend} for several
choices of $\omega$ at $\D=0.10$.  
With increasing $\omega$, the peak 
becomes broader and the $\eta_{x},\eta_{y}$-peaks 
develop more rapidly than that of the $\eta_{xy}$-peak. 
The location of each peak does not change appreciably up to 
$\omega \sim 0.2$ and shifts toward  $(\pi,\,\pi)$ at larger
$\omega$. This insensitivity to  $\omega$ is 
weakened for lower $\D$ and is limited to 
$\omega \slt 0.1$ at $\D =0.05$. 

\bfi[th]
\bc
 \includegraphics[width=8cm]{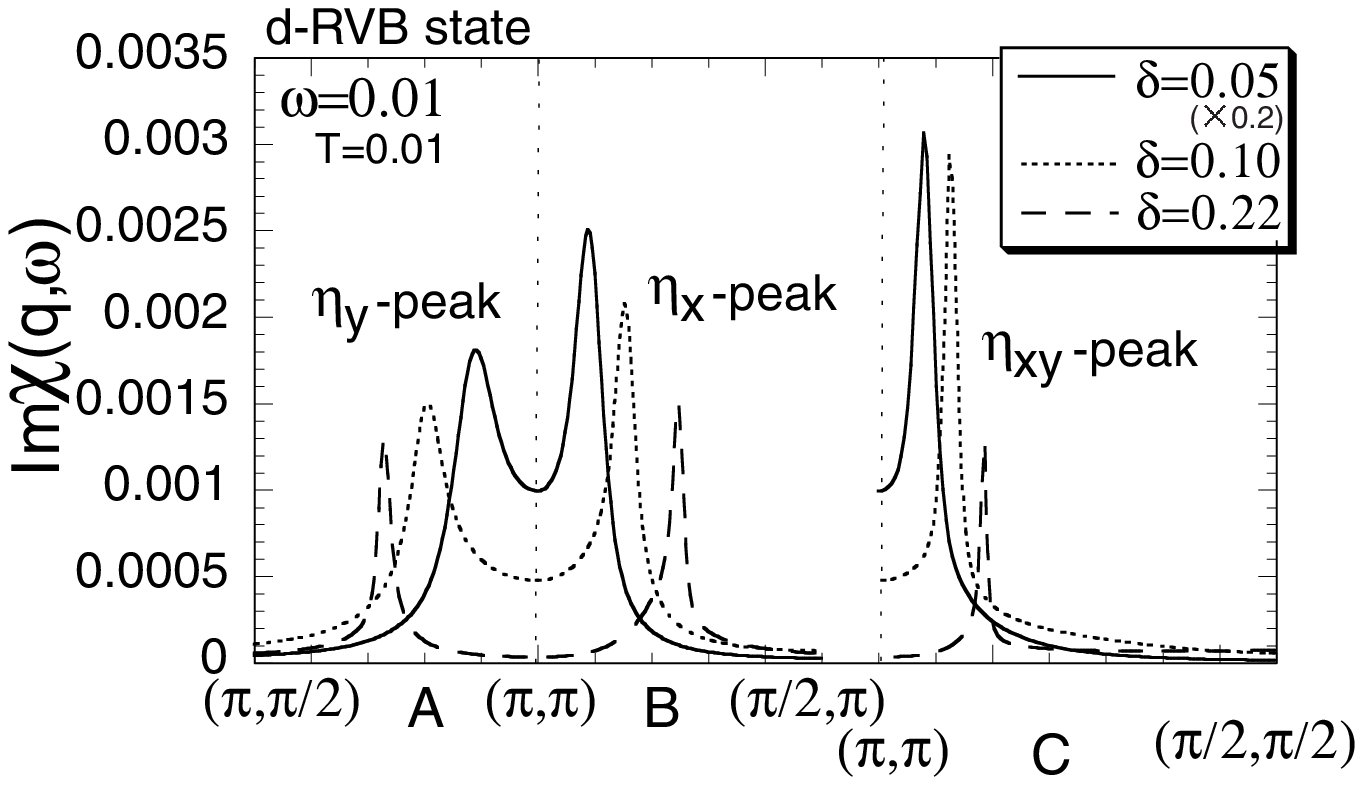}
     \caption{$\vq$-dependence of ${\rm Im} \chi(\vq,\,\omega)$ at
    $\omega=0.01$ for  several choices of 
$\D$ in the $d$-RVB state. The result for 
$\D=0.05$ is multiplied by 0.2.}
    \label{deltadepend}
\ec
\efi

In \fig{deltadepend}, we show the $\delta$-dependence of 
${\rm Im} \chi(\vq,\,\omega)$ at $\omega=0.01$. 
Both the $\eta_{x},\eta_{y}$-peaks and the $\eta_{xy}$-peak remain 
sharp down to low  $\delta$. 
The latter develops at lower $\D$
relative to the former.

\subsection{Effects of $d$-wave gap}  
To study effects of the $d$-wave gap, 
we show in \fig{wurvb} ${\rm Im} \chi(\vq,\,\omega)$ 
in the u-RVB state at several values of $\omega$. 
As in the $d$-RVB state (Fig.~\ref{wdepend}), 
sharp peaks exist up to $\omega \sim 0.1$, 
although such $\omega$-range is reduced compared with the $d$-RVB state,  
and the peak width gets broader. 
Hence, for the existence of the $\eta_{x},\eta_{y}$-peaks and the 
$\eta_{xy}$-peak, the $d$-wave gap is not essential but the q-1dFS is. 
(In \S 3.5, we will see that for the 2dFS 
such sharp peaks are possible only in the $d$-RVB state.)  

\bfi[ht]
\bc
 \includegraphics[width=8cm]{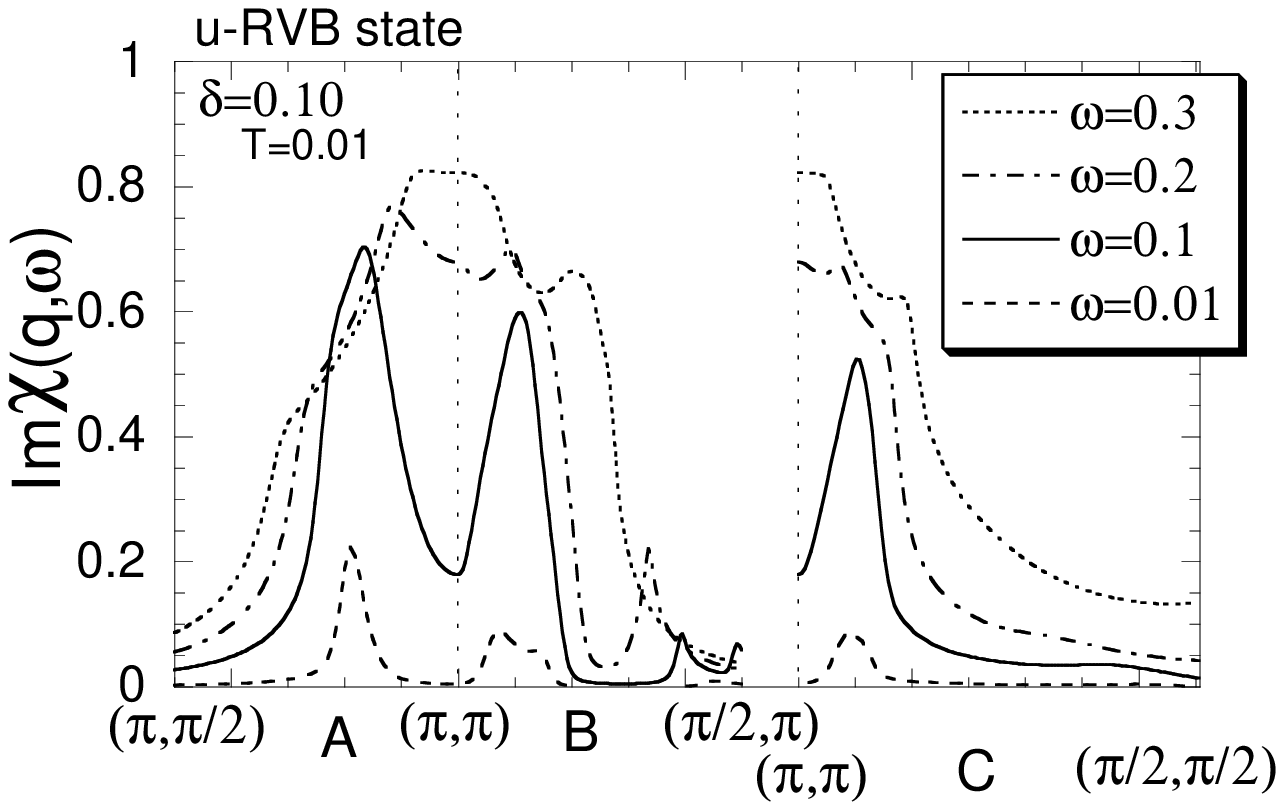}
     \caption{$\vq$-dependence of ${\rm Im} \chi(\vq,\,\omega)$ in the u-RVB
    state at several values of $\omega$.}
    \label{wurvb}
\ec
\efi

\bfi[ht]
\bc
 \includegraphics[width=8cm]{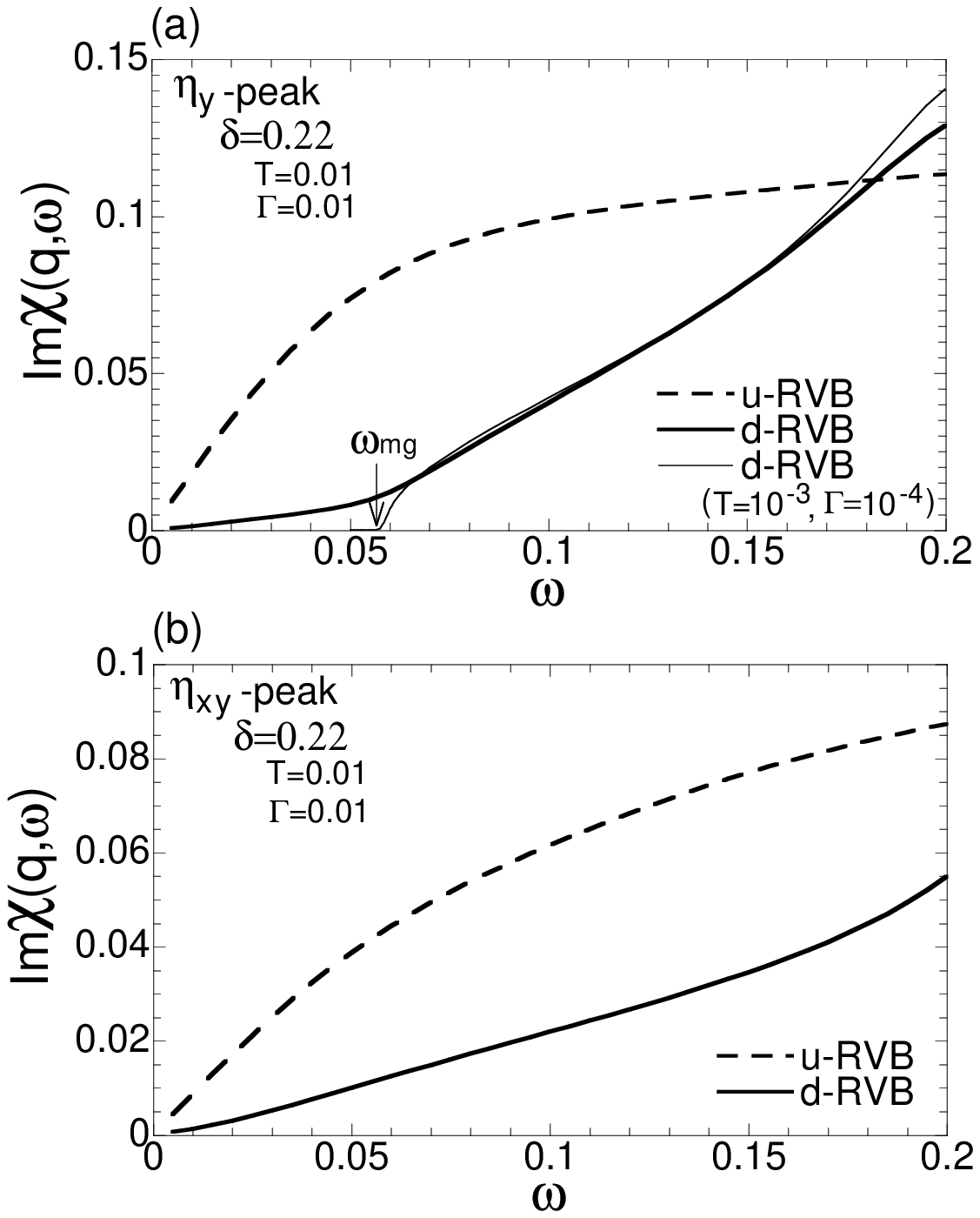}
     \caption{$\omega$-dependence of ${\rm Im} \chi(\vq,\,\omega)$ at $\vq$ 
     corresponding to the $\eta_{y}$-peak (a) and the
     $\eta_{xy}$-peak (b) in the $d$-RVB state (solid line) 
     and the u-RVB state (dashed line). 
     The thin solid line in (a)   is for lower temperature $T=10^{-3}J$ 
     and the smaller value of $\Gamma=10^{-4}J$. }
    \label{eta-wdepend}
\ec
\efi

\bfi[ht]
\bc
 \includegraphics[width=8cm]{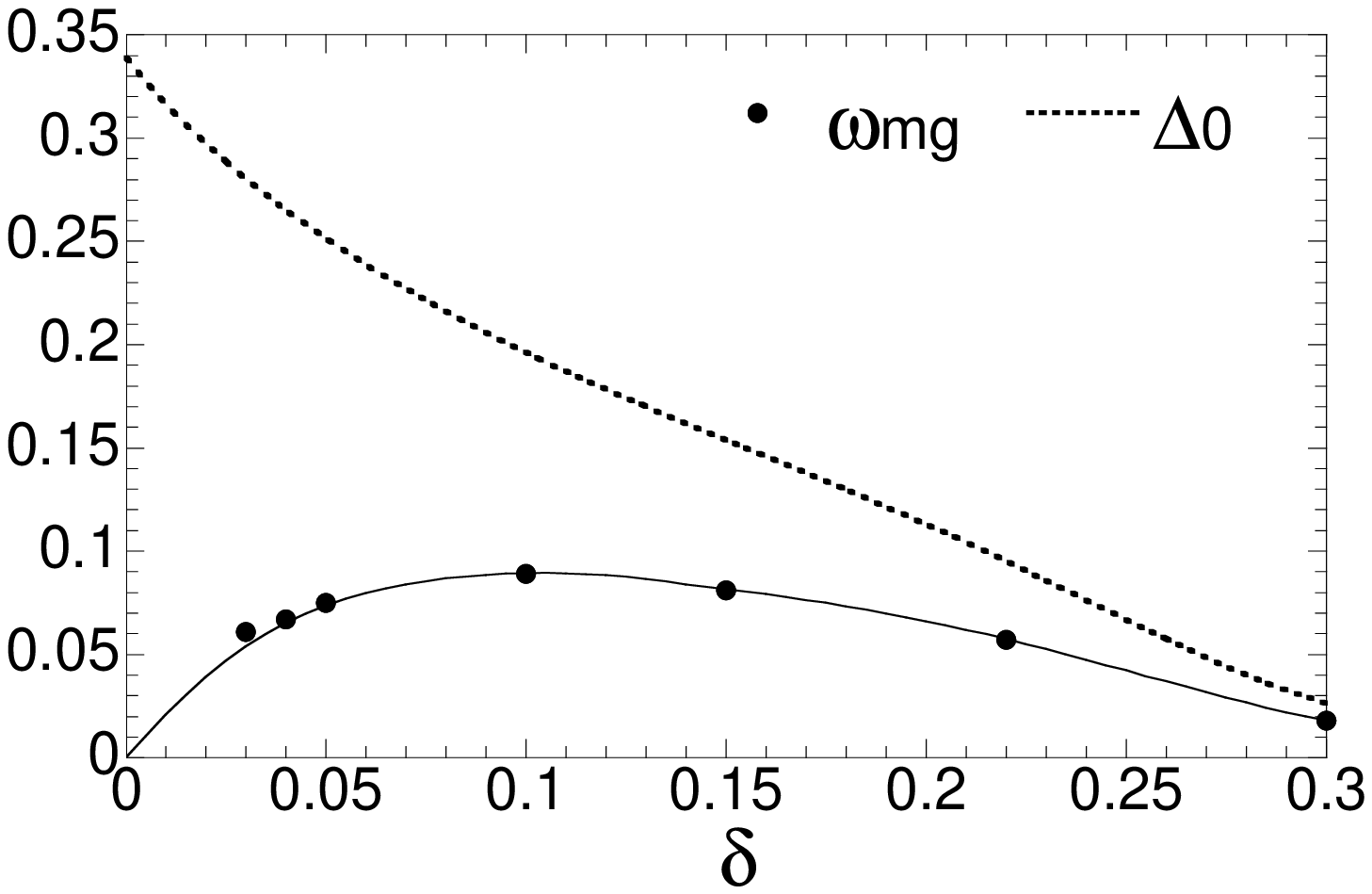}
     \caption{The magnetic gap 
     $\omega_{\rm mg}$ (filled circle) at the $\eta_{y}$-peak as a
    function of $\D$. 
    The solid line is drawn by using the analytic formulae,  
    eqs.~(\ref{wmg}), (\ref{etaxy}) and (\ref{etay}). 
    For comparison, the magnitude of the $d$-wave singlet order, 
     $\Delta_{0}$, is also shown (dotted line).}
    \label{omegamg}
\ec
\efi

\bfi[ht]
\bc
 \includegraphics[width=8cm]{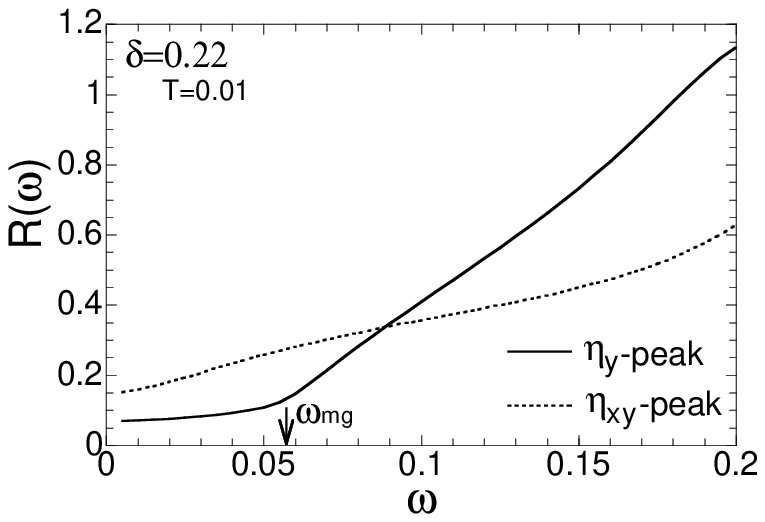}
     \caption{The degree of suppression of the $\eta_{y}$-peak and the
    $\eta_{xy}$-peak by the
    $d$-wave gap as a function of $\omega$. $R(\omega)$ is defined to be
    $\frac{{\rm Im}\chi(\vq,\,\omega)\left|_{d-{\rm RVB}}\right.}
{{\rm Im}\chi(\vq,\,\omega)\left|_{\rm u-RVB}\right.}$ with $\vq$
    corresponding to the peak position in each state. }
    \label{eta-wdepend2}
\ec
\efi

Figure \ref{eta-wdepend} shows ${\rm Im} \chi(\vq,\,\omega)$ as a function 
of $\omega$; $\vq$ is fixed to the peak position,  
$\eta_{y}=0.135\;(0.130)$ or $\eta_{xy}=0.066\; (0.058)$,  
in the $d$-RVB (u-RVB) state for $\delta=0.22$.  
Both the $\eta_{y}$-peak and the $\eta_{xy}$-peak
are suppressed by the $d$-wave gap,  
because the $d$-wave gap reduces the density of states responsible 
to the low energy scattering. 
For the $\eta_{y}$-peak in the $d$-RVB, 
a gap-like behavior appears 
at low $\omega$, and becomes 
clearer with decreasing both $T$ and $\Gamma$ as shown by the 
thin line ($T=10^{-3}$ and $\Gamma=10^{-4}$) in \fig{eta-wdepend}(a). 
This gap, which we call magnetic gap\cite{magneticgap} $\omega_{\rm mg}$,   
is approximately given by 
\be
\omega_{\rm mg}(\vq)=E_{\vq + \vk_{\rm node}}\; , \label{wmg} 
\ee
where $\vq$ is the scattering vector, $(\pi,\,\pi-2\pi\eta_{y})$, 
and $\vk_{\rm node}$ is a $d$-wave gap node position on the FS.  
Since the $\eta_{xy}$-peak in the $d$-RVB 
results mainly from the scattering between the $d$-wave gap nodes 
(\fig{nesting}), the value of $\vk_{\rm node}$ is estimated to be 
$- \frac{1}{2}(\pi-2\pi\eta_{xy},\,\pi-2\pi\eta_{xy})$. 
Substituting the values of $\eta_{xy}$ and $\eta_{y}$ into \eq{wmg}, 
we get $\omega_{\rm mg}(\vq)=0.057$, which is in good agreement with 
\fig{eta-wdepend}(a).  
(This good agreement has also been checked for other hole density.)  
In \fig{omegamg}, we plot $\omega_{\rm mg}(\vq)$ as a function of 
$\D$ together with the $d$-wave singlet order $\Delta_{0}$. 
Note the different $\D$-dependence between $\omega_{\rm mg}(\vq)$ and 
$\Delta_{0}$. 
(In Appendix~B, we give analytic formulae for 
estimating the values of $\eta_{xy}$ and $\eta_{y}$. 
The magnitude of the magnetic gap is then obtained from \eq{wmg} 
without any calculations of ${\rm Im} \chi(\vq,\,\omega)$.)

On the other hand, the magnetic gap is zero at  the 
$\eta_{xy}$-peak.
This does not, however, mean that effects of the 
$d$-wave gap are smaller compared with the $\eta_{y}$-peak. 
Using the result shown in \fig{eta-wdepend}, we show 
in \fig{eta-wdepend2} the ratio, 
\be
R(\omega)=\frac{{\rm Im} \chi(\vq,\,\omega)\left|_{d-{\rm RVB}}\right.}
{{\rm Im} \chi(\vq,\,\omega)\left|_{\rm u-RVB}\right.} \, ,
\ee 
for both the $\eta_{y}$-peak and the $\eta_{xy}$-peak 
as a function of $\omega$. 
At $\omega \approx 0$, the $d$-wave gap 
suppresses the $\eta_{y}$-peak more strongly than the $\eta_{xy}$-peak. 
However, once $\omega$ exceeds $\omega_{\rm mg}$, 
the suppression of the $\eta_{xy}$-peak becomes 
more prominent than the $\eta_{y}$-peak.

\subsection{Effects of interlayer hopping}
Next, we investigate effects of the interlayer hopping, 
$t_{\perp}$, on the (single-layer) results 
presented so far. This introduces mixing between the two kinds of q-1d
bands.  Details of the formalism are given in Appendix~A. 

Reflecting the relative shift  of Cu sites by 
$[\frac{1}{2},\frac{1}{2},\frac{1}{2}]$ (tetragonal notation) 
between the adjacent CuO$_{2}$ layers, the band dispersion along the 
$k_{z}$-direction is obtained as 
\be
\epsilon_{\vk}=- 8 t_{\perp}\delta \cos \frac{k_{x}}{2} \cos \frac{k_{y}}{2}
\cos \frac{k_{z}}{2} \; .
\ee
We set the interlayer transfer integral to be 
$t_{\perp}=0.05 t^{(1)}$ so that the band width of $\epsilon_{\vk}$ 
is about 0.1 times that of $\xi_{\vk}$\cite{pickett}. 
(We neglect the interlayer magnetic interaction, whose order is 
$\sim 10^{-5}J$\cite{thio,peters,lyons}.) 
In \fig{q1dFSs}, we show the FS at $k_{z}=0$ and $\pi$ obtained  
by setting $\Delta_{\vk}=0$ in \eq{tzE}. The FS consists of the 
outer FS (solid line in \fig{q1dFSs}) and the inner FS (gray line).  

\bfi[ht]
\bc
 \includegraphics[width=8cm]{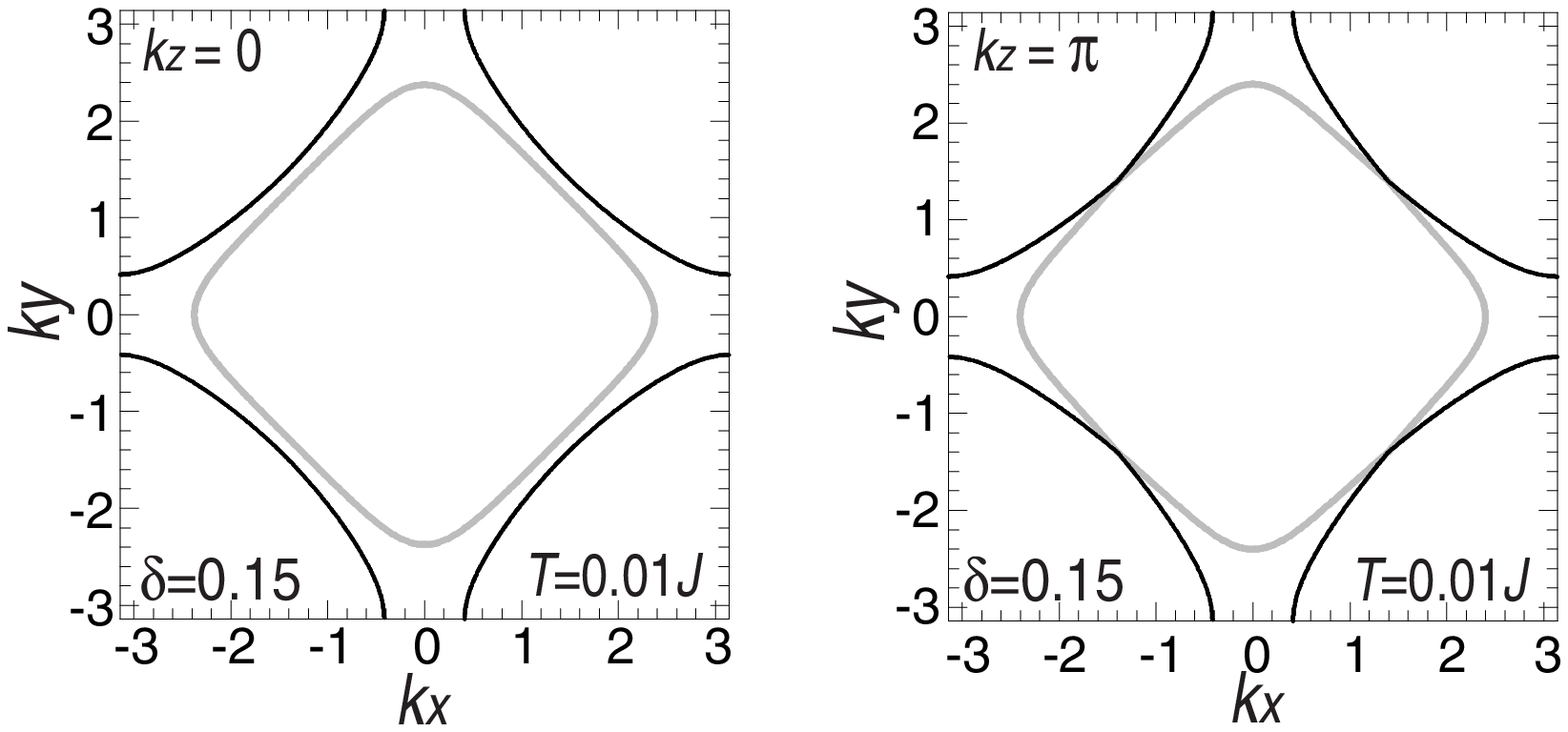}
     \caption{Theoretical FS at $k_{z}=0$ and $\pi$ at low $T$ in
    our quasi-one-dimensional picture of the FS 
    in the presence of interlayer
    hopping.  (see also \fig{q1dpicture}.) 
    The FS consists of the outer FS (solid line) and 
    the inner FS (gray line).}
    \label{q1dFSs}
\ec
\efi

Taking $q_{z}=0$, we show in Figs.~\ref{tz}(a) and \ref{tz}(b)    
the $\vq$-dependence of ${\rm Im} \chi(\vq,\,\omega)$ 
at $\omega=0.01$ for $\delta=0.10$ and $0.22$ (bold solid line) in the 
$d$-RVB state, comparing it with the (single-layer)    
result for the q-1dFS($x$) (thin dotted line). 
The IC-peaks at  
$(\pi - 2\pi \eta_{_{\rm IC}},\,\pi)$ and 
$(\pi,\,\pi - 2\pi \eta_{_{\rm IC}})$ remain sharp, 
and recover the fourfold symmetry around $(\pi,\,\pi)$.    
The peak position is almost the same as that of the 
(single-layer) $\eta_{x},\eta_{y}$-peaks, 
and the width becomes slightly broader.
On the other hand, the DIC-peak at 
$(\pi- 2\pi \eta_{_{\rm DIC}},\,\pi - 2\pi \eta_{_{\rm DIC}})$  
gets much broader compared with the (single-layer) $\eta_{xy}$-peak 
and is largely suppressed at high $\D$ $(\sgt 0.22)$. 

\bfi[ht]
\bc
 \includegraphics[width=13cm]{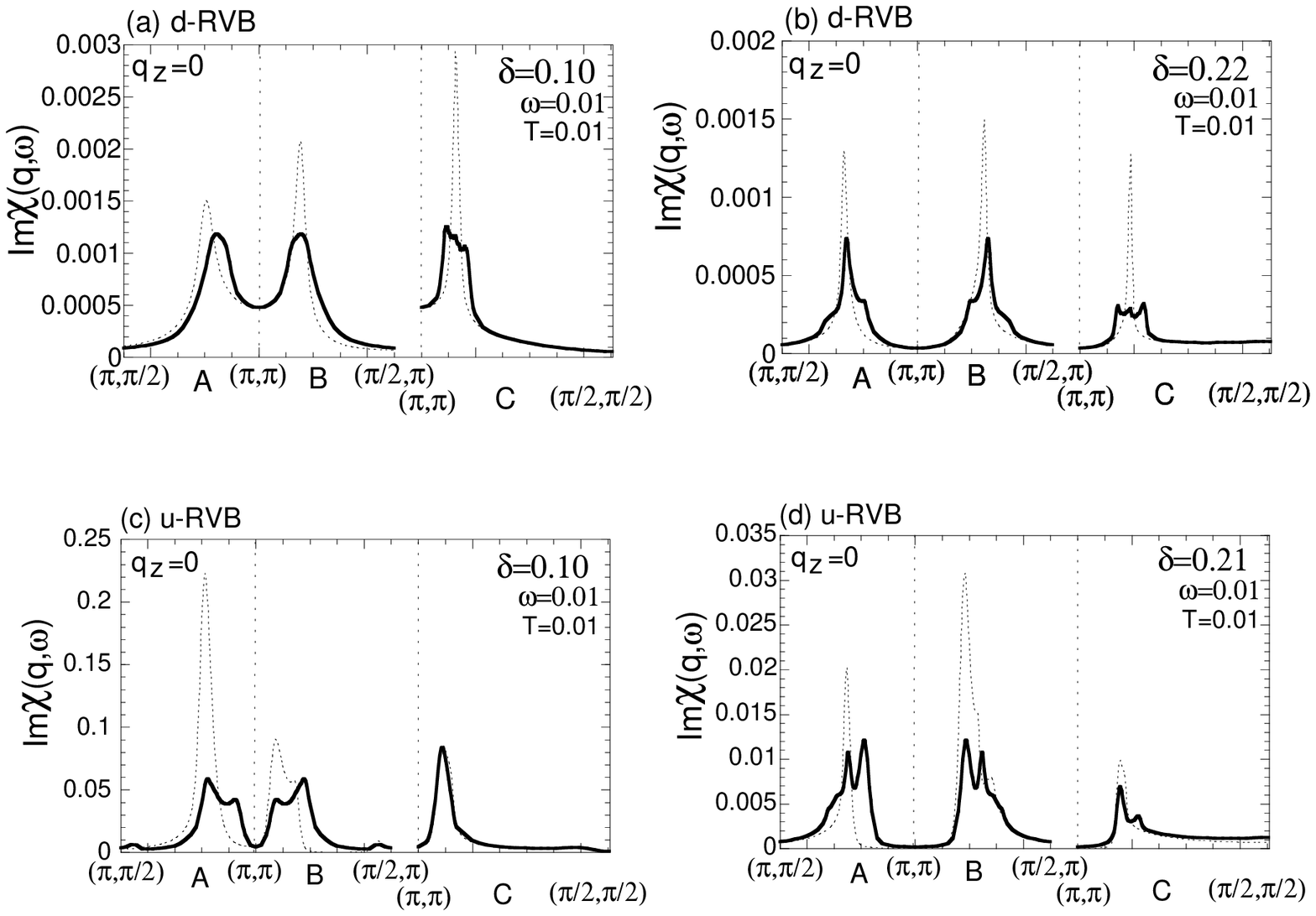}
     \caption{$\vq$-dependence of ${\rm Im} \chi(\vq,\,\omega)$ at
    $\omega=0.01$ and $q_{z}=0$  
    in the presence of interlayer hopping (solid lines). 
    It is plotted in both the $d$-RVB state and the u-RVB state for 
    several choices of $\D$. 
    The rough topped DIC-peak in (a) and (b) will be due to the artifact 
    of the present calculation 
    and should be interpreted as a smooth one. (see the last paragraph  in 
    Appendix~A.)  
    The (single-layer) results for the q-1dFS($x$) are also
    plotted for comparison (dotted lines).}
    \label{tz}
\ec
\efi

Figures \ref{tz}(c) and \ref{tz}(d) 
show the results in the u-RVB state. As in the   
$d$-RVB, IC-peaks are seen, but 
the line shape changes qualitatively. 
The IC-peak has a hump on  the $(\pi,\,\pi)$ side 
at  $\delta =0.10$. This hump 
originates from the (single-layer) $\eta_{x}$-peak for the q-1dFS($x$).  
Such structure develops into a double-peak structure at higher $\D$. 
With further increasing $\D$ $(\sgt 0.23)$, the double 
peaks merge into a single sharp peak 
because of the decrease of the band anisotropy,  
$\alpha \rightarrow 1$. 
As for the DIC-peak, the line shape is almost the same as that 
for the (single-layer) q-1dFS($x$).  
The peak height relative to the IC-peak 
is suppressed at higher $\D$ as in the case of the $d$-RVB. 

With increasing $\omega$, the IC-peak develops more rapidly 
than the DIC-peak, 
and become broader so that the fine structures such as the hump or 
the double-peak structure are smeared. 
These $\omega$-dependence share the common features to the
(single-layer) results shown in Figs.~\ref{wdepend} and \ref{wurvb}.

In \fig{etadelta}, we plot the $\delta$-dependence of $\eta_{_{\rm IC}}$ 
and $\eta_{_{\rm DIC}}$ at $\omega=0.01{\rm }$ 
in the $d$-RVB (a) and u-RVB (b) 
--- the $\eta_{_{\rm DIC}}$ at high $\D$ is not shown 
since the peak height of  the DIC-peak 
is less than 50\% of the IC-peak.     
(For the low doping region $\delta \leq 0.05$ in the u-RVB, 
we take $r=0.2$, instead of $r=0.35$, to avoid 
magnetic instability in \eq{RPA}.)   
In both states, the values of $\eta_{_{\rm IC}}$ and 
$\eta_{_{\rm DIC}}$  increase smoothly as a function of $\D$, 
except in the region $0.18 \slt \D \slt 0.23$ in the u-RVB, where 
${\rm Im} \chi(\vq,\,\omega)$ shows the double peaks. 
Their peak positions are 
plotted by filled circles with  different size so that 
the smaller indicates the position of the lower peak. 
Since the relative height changes from the larger $\eta_{_{\rm IC}}$ 
to the smaller $\eta_{_{\rm IC}}$ in $0.19\slt \D\slt 0.21$ in
\fig{etadelta}(b), 
this may appear as a saturation behavior of $\eta_{_{\rm IC}}$  
in $0.15 \slt \D \slt  0.20$.  

\bfi[ht]
\bc
 \includegraphics[width=8cm]{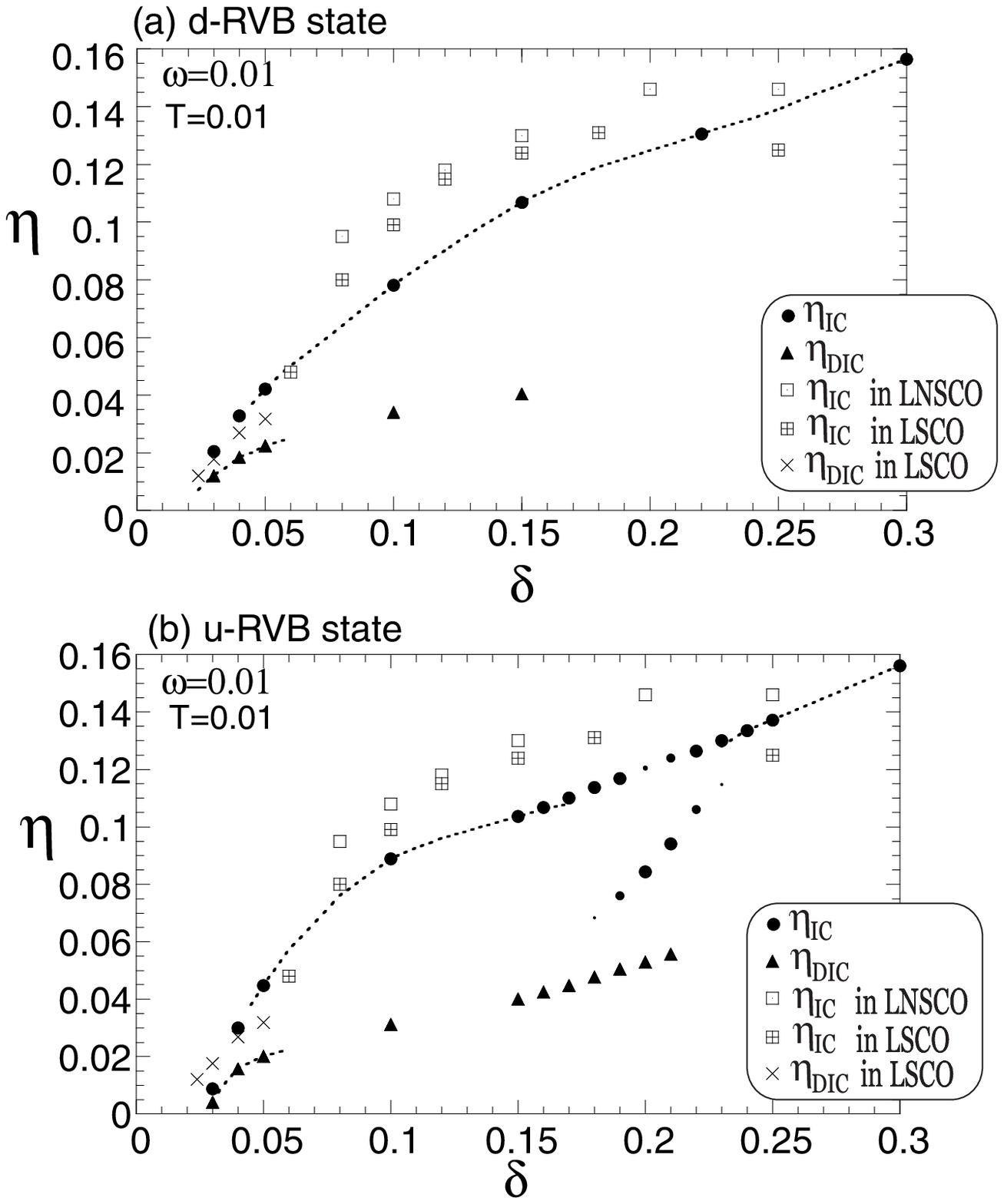}
     \caption{Incommensurability, $\eta_{_{\rm IC}}$ and
    $\eta_{_{\rm DIC}}$, as a function of $\D$ for the $d$-RVB state
    (a) and the u-RVB state (b). In (b), two filled circles at each
    $\delta$ in the range $0.18 \le \D \le 0.23$ correspond to 
    the double-peak positions with the higher peak being indicated by the 
larger circle; 
    the relative difference of the peak height is about 
    $18\% (\D=0.18)$, $2\% (\D=0.19)$, $13\% (\D=0.20)$, $10\%
    (\D=0.21)$, $7\% (\D=0.22)$, $17\% (\D=0.23)$, respectively. 
    Experimental data in LNSCO\cite{ichikawaPD} 
    and in LSCO\cite{yamada,matsuda2} are also plotted; 
    note that the IC-peak has
    been reported in  $\D \ge 0.06$ and the DIC-peak in $\D \le
    0.05$. The dotted lines are drawn for  guides to the eye for 
    comparison with experimental data.  }
    \label{etadelta}
\ec
\efi

At higher $\omega$, the $\D$-dependence of $\eta_{_{\rm IC}}$ and 
$\eta_{_{\rm DIC}}$ in the $d$-RVB is almost the same as that at
$\omega=0.01$, since the positions of the (single-layer)
$\eta_{x},\eta_{y},\eta_{xy}$-peaks do not change appreciably with
$\omega$ as shown in \fig{wdepend}. 
In the u-RVB state, however, the value of $\eta_{_{\rm
IC}}$ gets smaller especially in low $\D(\slt 0.15)$ and possible
saturation behavior in $0.15 \slt \D \slt 0.20$ is smeared out. 
This qualitative difference from that at $\omega=0.01$ is due to the 
smearing of the hump and the double-peak structures 
of ${\rm Im} \chi(\vq,\,\omega)$.  
In contrast, 
the value of $\eta_{_{\rm DIC}}$ remains almost the same with increasing  
$\omega$, as seen in the (single-layer) result shown in \fig{wurvb}.

\subsection{Temperature dependence of ${\rm Im} \chi(\protect\vq,\,\omega)$}
So far we have seen magnetic excitation at low $T$ $(=0.01 J)$. With 
increasing $T$, the following two effects may be expected.  
(i) The IC-peaks are thermally smeared out to form a C-peak. 
(ii) The q-1dFS is destabilized and is replaced by a \lq 2dFS'.  
Here the possible \lq 2dFS' is the one obtained with the same 
band parameters as used in \S~2.1,   
but 
$F_{x}=F_{y}$ is kept for each $\delta$.  
The resulting FS at $\D=0.15$ is \lq electron-like' centered 
at $(0,\,0)$ as shown in \fig{previousFS}. 

\bfi[ht]
\bc
 \includegraphics[width=4cm]{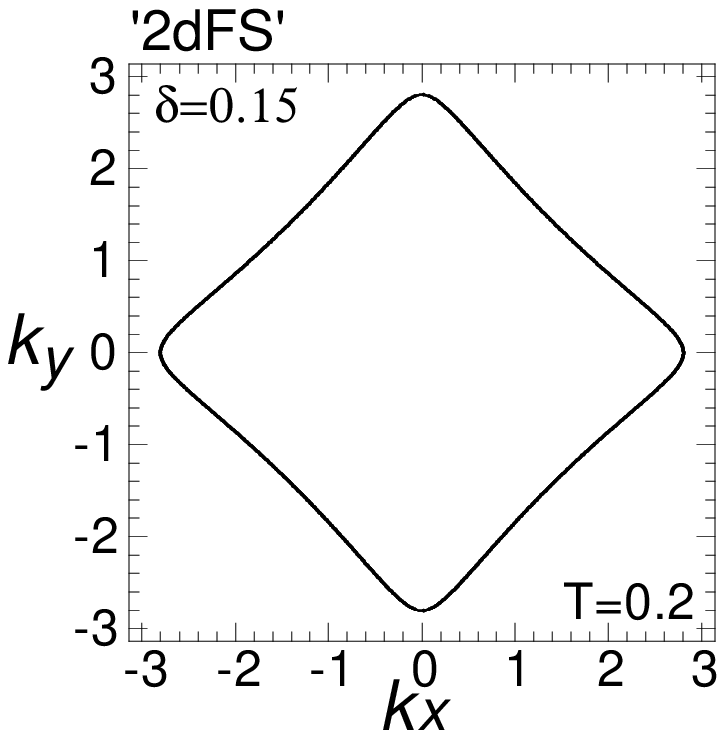}
     \caption{The \lq 2dFS', which may be 
     realized at high $T$ in the present 
    quasi-1d picture of the FS.} 
    \label{previousFS}
\ec
\efi

We thus calculate ${\rm Im} \chi(\vq,\,\omega)$ for several choices of 
$T$ for both 
the q-1dFS and the \lq 2dFS' in the u-RVB state. 
We include the interlayer hopping and 
set $\omega=0.1$ and $\D=0.15$. 
Figure~\ref{tdepend} shows that the IC-peak exists for both FSs in 
$T \slt 0.1$.  At $T=0.2$, however, 
a broad C-peak is realized for the q-1dFS,  
and a weak IC-peak or an essentially  flat 
topped commensurate (C-) peak for the \lq 2dFS'. 

\bfi[ht]
\bc
 \includegraphics[width=13cm]{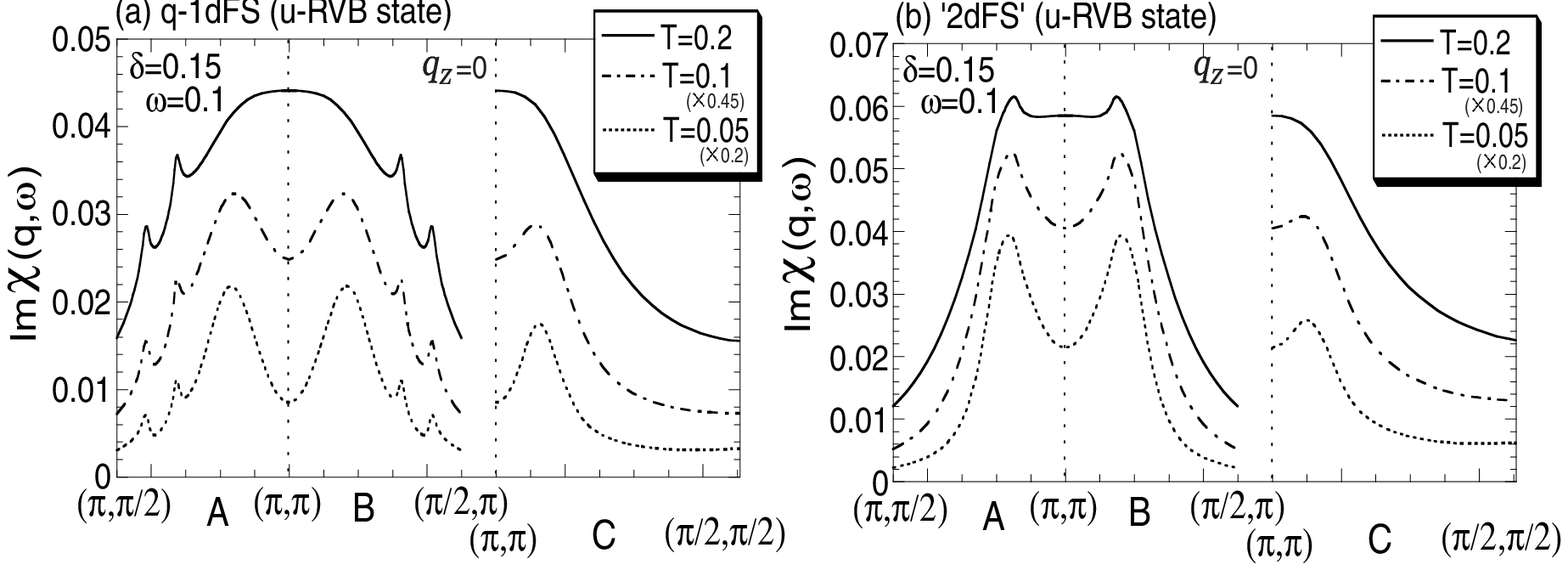}
     \caption{$\vq$-dependence of ${\rm Im} \chi(\vq,\,\omega)$ for  several 
choices of $T$
    for the q-1dFS (a) and the \lq 2dFS' (b). Note that the 
    interlayer hopping is included. The data at $T=0.1$
    and $0.05$ are multiplied by 0.45 and 0.2, respectively. }
    \label{tdepend}
\ec
\efi

\subsection{Comparison with 2dFS} 
As we discussed in the previous paper\cite{yamase1}, the present 
q-1d picture of the FS 
is consistent with the ARPES data in LSCO\cite{ino}. 
Moreover, recent data in ARPES\cite{zhou2}, the $\vk$-space 
distribution of low-energy spectral weight integrated down 
to 30meV below Fermi energy, has 
turned out to be consistent with our predicted FS shown in
\fig{q1dFSs}. We, however, note a different arguments from ours  
that the FS in LSCO will be the 2dFS shown  in
\fig{2dFS}\cite{ino}. (Note the difference in topology from 
the \lq 2dFS' shown in \fig{previousFS}.) 
We thus investigate the difference in 
magnetic excitation between the q-1dFS and the 2dFS here. 

\bfi[bh]
\bc
 \includegraphics[width=4cm]{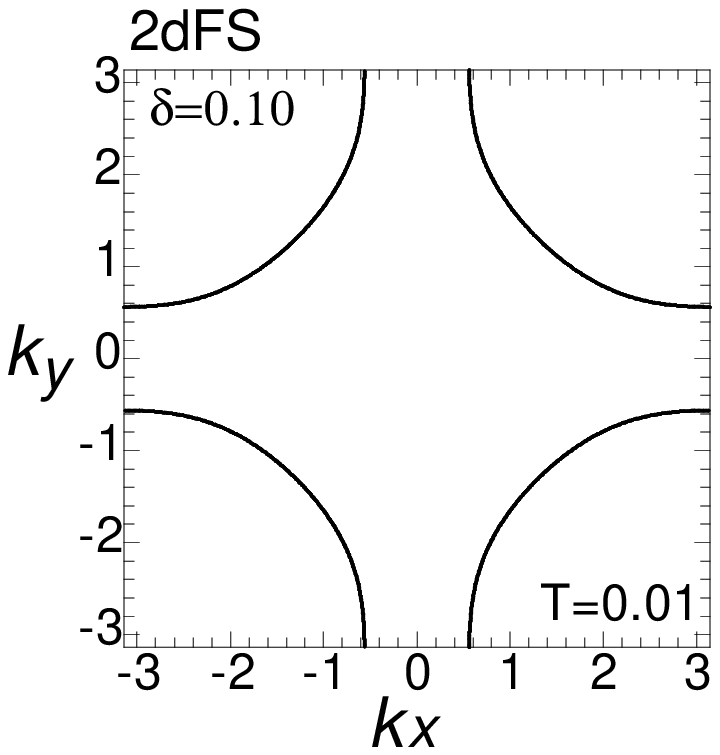}
     \caption{The 2dFS discussed in the ARPES study for LSCO with 
     $\D \slt 0.20$\cite{ino}. This FS is 
     different from the present 
     quasi-one-dimensional picture. 
     (see Figs.~\ref{q1dpicture} and \ref{q1dFSs}.)}
    \label{2dFS}
\ec
\efi

We take the band parameters, $t^{(1)}/J=4, t^{(2)}/t^{(1)}=-1/6$, and 
$t^{(3)}/t^{(1)}=1/5$ and determine mean fields by minimizing the free 
energy in the same fashion as \S 2.1, 
but now with $F_{x}=F_{y}$. We then obtain 2dFS shown in \fig{2dFS}.

Using this 2dFS, we show ${\rm Im} \chi(\vq,\,\omega)$ 
for several choices of $\omega$ 
in Figs.~\ref{2dwdepend}(a) and \ref{2dwdepend}(b). 
In the $d$-RVB, both the sharp IC-peak and the sharp DIC-peak are
realized. 
In the u-RVB, however, 
a broad  C-peak 
becomes dominant and the remnants  of the IC,DIC-peaks appear as
weak substructures. 
This feature contrasts with the case of the q-1dFS  
where the sharp IC,DIC-peaks are realized even in the u-RVB. 
This difference may be understood by noting that 
the 2dFS allows the scattering vector $(\pi,\,\pi)$ from 
one point to the other on the 
FS while the q-1dFS does not. 
(When we introduce hybridization between 
the two kinds of the q-1d bands, the resulting 
FS (\fig{q1dFSs}) allows such $(\pi,\,\pi)$ scattering vectors. 
However, these scatterings occur through the interlayer
hopping and do not contribute appreciably 
to ${\rm Im} \chi(\vq,\,\omega)$ as seen from Figs.~\ref{tz}(c) and \ref{tz}(d).)

\bfi[bh]
\bc
 \includegraphics[width=8cm]{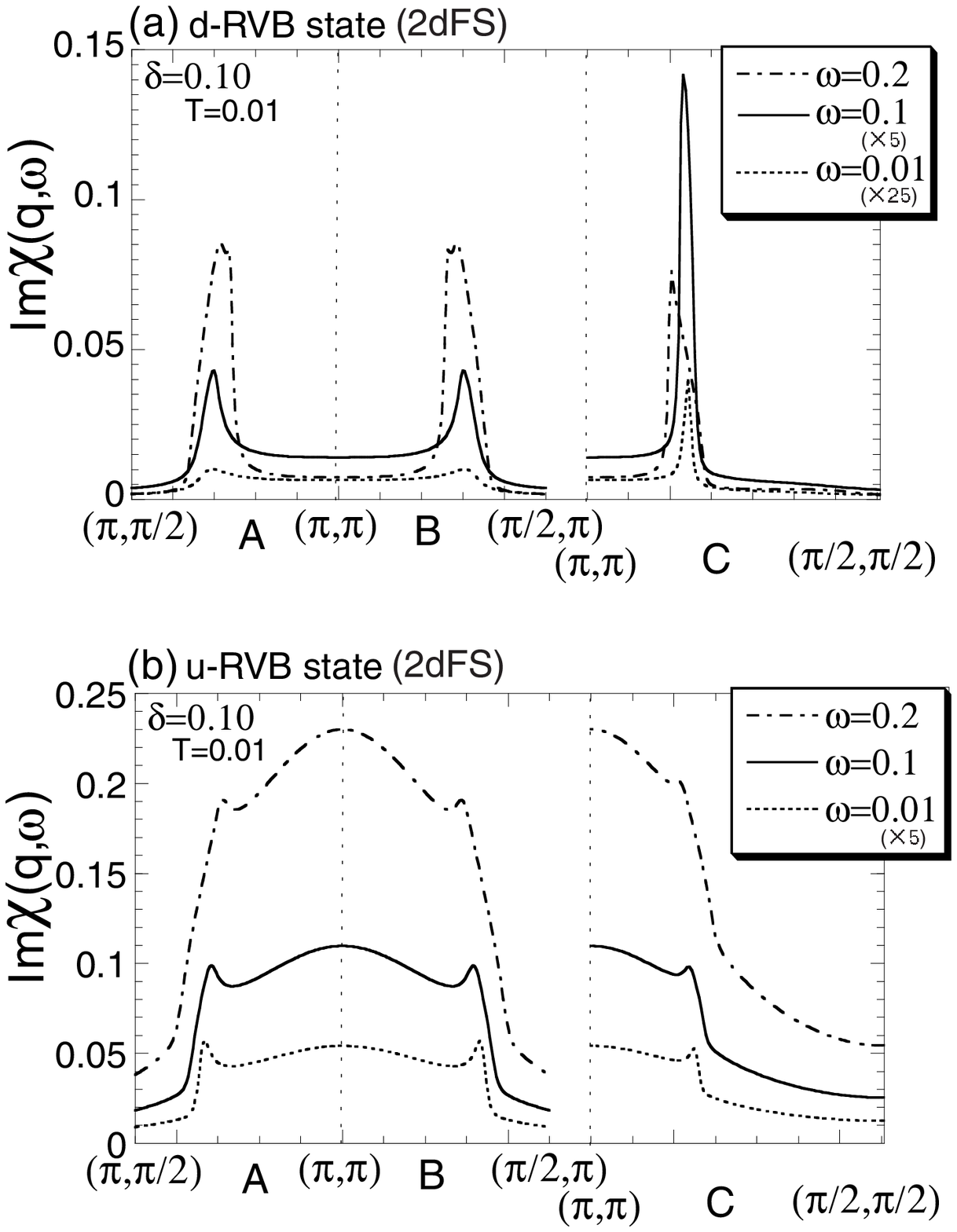}
     \caption{$\vq$-dependence of ${\rm Im} \chi(\vq,\,\omega)$ for several
    choices of $\omega$ for the 2dFS 
    in the $d$-RVB state (a) and the u-RVB
    state (b). (The interlayer hopping is not included.) 
    The data at $\omega=0.1$ and $0.01$ in (a), and 
    the data at $\omega=0.01$ in (b) are multiplied by 5, 25 and
    5, respectively.}
    \label{2dwdepend}
\ec
\efi

\section{Discussion}
\subsection{Possible q-1d picture of FS in LSCO systems}
Now we discuss a possible  q-1d picture of the FS in  LSCO systems  
from the viewpoint of magnetic excitation.  
We take the following four subjects: (i) IC-peak at high temperature, 
(ii) $\omega$-dependence and magnetic gap, 
(iii) incommensurability versus
hole density, and (iv) IC-peak versus  DIC-peak. 

\subsubsection{IC-peak at high Temperature}
Experimentally, the IC-peak has been observed 
at $T=80{\rm K}$ for LSCO with 
$\delta=0.14$\cite{aeppli}, and 
the spin gap behavior has not been observed at
least above $T\approx 80$K\cite{mason,nakano,momono,ohsugi}.  
Thus it is difficult to understand such experimental data 
in terms of the 2dFS, since as shown in \fig{2dwdepend}, 
the IC-peak     
for the 2dFS is realized only in the $d$-RVB state (with the spin gap), 
and is replaced by the essentially C-peak in the u-RVB state 
(without the spin gap).  
Rather, such experimental data is consistent with the results based on
the q-1d picture of the FS as shown in Figs.~\ref{wurvb}, \ref{tz}(c), 
\ref{tz}(d), \ref{tdepend}(a) and \ref{tdepend}(b), 
where IC-peaks have been realized in the u-RVB state
at least for $T\slt 0.1J$; the value of $J$ is estimated as 
$\sim 135$meV\cite{lyons}.

The data in ref.~\onlinecite{aeppli} also indicates  
a weak IC-peak (or a possible broad C-peak) at $T=297{\rm K}$, 
where the lattice structure is the high-temperature tetragonal (HTT).  
Our obtained results at $T=0.2J$ 
in Figs.~\ref{tdepend}(a) and \ref{tdepend}(b) do not 
contradict with this experimental data for either FS, 
although  we expect the realization of a \lq 2dFS' at $T=297$K, since,  
according to our previous arguments\cite{yamase2} (see \S1),  
a q-1dFS may be realized in the presence of the LTT structure or 
its fluctuation,

\subsubsection{$\omega$-dependence and magnetic gap}
The (single-layer) results, Figs.~\ref{wdepend} and \ref{wurvb},  
have indicated that the positions of the sharp 
IC,DIC-peaks do not depend  appreciably on $\omega$ 
up to $\omega \approx  0.1$--$0.2J$. 
Thus our results, including the 
energy scale,  are consistent with experiments\cite{matsuda1,mason,chlee00}. 
One may recall a discussion in \S~3.3 
that in the presence of interlayer hopping, 
the precise values of $\eta_{_{\rm IC}}$ in the u-RVB have been
sensitive to $\omega$ especially for low $\D (\slt 0.15)$. 
This sensitivity, however, has resulted from the smearing of 
a hump structure of IC-peak with increasing $\omega$, 
and the degree of change of $\eta_{_{\rm IC}}$ is limited within 
the peak width. Such change may not be essential in comparison with 
experiments. 

As shown  in \fig{eta-wdepend2}, with increasing $\omega$, 
the degree of suppression by the $d$-wave
gap is substantially weakened at the $\eta_{y}$-peak more than at the
$\eta_{xy}$-peak.  
This is consistent with data by Lake
{\it et al.}\cite{lake}. The argument in ref.~\onlinecite{lake} that 
the value of $\omega_{\rm mg}(\vq)$ is momentum-{\it independent},
however,     
can not be understood within the present result  
where $\omega_{\rm mg} \ne 0$ at the IC-peak while $\omega_{\rm mg} = 0$ 
at the DIC-peak. 

As for the value of $\omega_{\rm mg}$ at the IC-peak,  
it has been reported that 
$\omega_{\rm  mg}=6$--$7$ meV 
at  $\D=0.15, 0.16, 0.18$\cite{chlee00,lake} and $\omega_{\rm  mg}=0$        
at $\D=0.10, 0.25$\cite{chlee00}. 
The former data is consistent with the 
present result semiquantitatively as shown in \fig{omegamg}, 
but the latter is not. 
For the data $\omega_{\rm mg}=0$ at $\D=0.25$, 
we note the experimental fact\cite{yamada} that 
the hole density is close to  the phase boundary between 
the superconducting state and the metallic state. 
Because of such proximity to the normal state, 
(i) the value of $\omega_{\rm mg}$ may be smaller than 
the present estimation (\fig{omegamg}), and it may be difficult 
to observe such small $\omega_{\rm mg}$, 
and (ii) 
the fermion damping constant $\Gamma$ may be 
larger than that in $\D=0.15$-$0.18$, which will  
smear the clear magnetic gap behavior as demonstrated 
in \fig{eta-wdepend}(a). 
To understand the data at $\D=0.10$, 
further detailed theoretical studies\cite{yamase4} 
are required as  to 
why static IC-AF order is stabilized at 
$\D=0.10$\cite{matsushita},
0.12\cite{suzuki,kimura} and  0.13\cite{matsushita}, 
since these static order will  
enhance the spectral weight near $\omega \approx 0$ meV to 
smear the magnetic gap.

\subsubsection{Incommensurability versus hole density}
Considering the experimental indication   
that the value of 
incommensurability observed by the {\it elastic} neutron scattering 
is almost the same with that  by the {\it inelastic} one,  
we note the following experimental data: 
$\eta_{_{\rm IC}}$ observed by the {\it inelastic} neutron 
scattering in LSCO\cite{yamada,yamada3} at $T\approx T_{\rm c}$   
and by the {\it elastic} scattering 
in LNSCO\cite{ichikawaPD,ichikawa2}  
below $T_{\rm c}$,  and $\eta_{_{\rm DIC}}$ observed by the {\it elastic} 
scattering in LSCO\cite{matsuda2,wakimoto} at $T>T_{\rm c}=0$. 
Since our theory predicts the qualitatively different $\D$-dependence 
of $\eta_{_{\rm IC}}$ and $\eta_{_{\rm DIC}}$ between in the 
$d$-RVB state and in the u-RVB state (\fig{etadelta}), 
the experimental data should be 
compared in either state.  It is,  however, not obvious which state 
should be taken. For the data taken at $T \sgt T_{\rm c}$, 
there is a controversial issue whether 
the spin gap exists 
in LSCO\cite{fukuyama,momono,nakano,mason,chlee00,ohsugi}. 
For the elastic data below $T_{\rm c}$, it might be  reasonable 
to take the $d$-RVB state. However, 
the magnetic gap at the IC-peak in the $d$-RVB state 
is finite 
in the present theory 
while elastic data indicate that it is zero; 
we may not be limited to the  $d$-RVB state at present.  
Thus, leaving these to  future problems, 
we here make comparison with experimental 
data by taking both the $d$-RVB state and the u-RVB state. 

In Figs.~\ref{etadelta}(a) and \ref{etadelta}(b),  
we focus on the DIC-peak for 
\mbox{$0.02 \slt \delta \slt 0.05$} 
and the IC-peak for \mbox{$0.05\slt$}   
\mbox{$\delta \slt 0.30$}, although   
we have both the IC-peak and the DIC-peak 
at each $\D$ in the present study. 
We see that for both state, the values of $\eta_{_{\rm IC}}$ and
$\eta_{_{\rm DIC}}$ are somewhat smaller than  
the experimental values, 
but the semiquantitative agreement is obtained. 
More quantitative agreement may be obtained by the fine 
tuning of the FS, since as seen from \fig{nesting}, the value of  
incommensurability is sensitive to the position of the FS near 
$(\pi/2,\,\pi/2)$. 
In fact, we find that the FS used in the present analysis 
(\fig{q1dFSs}) lies somewhat outer near $(\pi/2,\,\pi/2)$  compared 
with  the position  of the observed low-energy spectral 
weight\cite{zhou2}, although we have fitted the FS 
near $(\pi,\,0)$ and $(0,\,\pi)$ 
to the observed FS segments\cite{ino}.  

The experimental data in LSCO by Yamada {\it et al.}\cite{yamada}
show that 
the values of $\eta_{_{\rm IC}}$ saturate in $\D \sgt 0.15$. 
We have obtained similar possible saturation behavior in $0.15 \slt \D
\slt 0.20$ in the u-RVB state for $\omega=0.01$ (\fig{etadelta}(b)).  
As discussed in \S~3.3, this possible saturation, however, 
has resulted from the fine structures of 
${\rm Im} \chi(\vq,\,\omega)$   
and is easily smeared out with increasing $\omega$. 
It is beyond the present calculation to associate such subtle structures  
with the experimental data. 
We leave to a future study why the value of $\eta_{_{\rm IC}}$
saturates at high $\D$ in LSCO.

\subsubsection{IC-peak versus DIC-peak}
We have obtained both the IC-peak and the DIC-peak at each $\D$. 
Which peak should be observed experimentally? 
As shown in Figs.~\ref{tz}(b) and \ref{tz}(d), 
the DIC-peak is substantially suppressed 
at high $\delta$, which indicates  that 
the IC-peak becomes dominant for high $\D$ $(\sgt 0.20)$. 
This is consistent with experiments\cite{yamada,ichikawaPD}. 
For lower $\D$, however, we expect both the IC-peak and the DIC-peak,  
and the former develops more rapidly than the latter with increasing 
$\omega$. 
At present, we have no definite answer to the question 
why the IC-peak has been observed only in 
$\D \sgt 0.05$\cite{yamada,ichikawaPD} and replaced 
by the DIC-peak 
in $0.02 \slt \delta \slt 0.05$\cite{wakimoto,matsuda2,fujita01}.

\subsection{Magnetic excitation in YBCO systems} 
We have seen  in \S 3.4 that for the 2dFS (\fig{2dFS})  
the IC-peak is realized only in the $d$-RVB state and is 
replaced by the essentially C-peak in the 
u-RVB state within the $\omega$-range studied here.  
This feature is consistent with experimental 
data for  YBCO\cite{dai,arai,bourges}. 
Moreover, the 2dFS shown in \fig{2dFS} is 
consistent with the FS observed by ARPES\cite{schabel}. 
We therefore consider that magnetic excitation in YBCO
systems may be understood on the basis of the 2dFS, 
as has been discussed theoretically\cite{tanamoto2,lee,levin}.

\subsection{Relation to \lq spin-charge stripes' hypothesis}
We have seen that fermiology can be a central concept for 
understanding magnetic excitation in high-$T_{\rm c}$ cuprates. 
This viewpoint contrasts with the \lq spin-charge stripes' 
scenario\cite{tranquada1,tranquada2,tranquada3} where it is the
formation of \lq charge stripes', not effects of the FS, that gives 
rise to the magnetic IC,DIC-peaks. 

Nonetheless, possible realization of the \lq charge stripes' 
is interesting. 
In the present study, we have assumed that 
the charge density is uniform. When we relax this restriction,  
some kind of charge ordering may be stabilized 
in the state with a q-1dFS, which we are trying to clarify. 

\subsection{Degree of band anisotropy at high doping rate} 
In \S2.1, we have fit a q-1dFS near $(\pi,\, 0)$ or $(0,\, \pi)$ 
to the observed FS segments by ARPES\cite{ino}. 
Figure~\ref{alpha-delta} 
implies that band anisotropy remains even at high $\D \, (\sgt 0.20)$. Such 
band anisotropy, however, can not be understood in terms of 
our previous arguments\cite{yamase2} (see \S1), 
since the crystal structure in LSCO with $\D \sgt 0.20$\cite{yamada} is 
the HTT where 
we expect a \lq 2dFS'. 
This problem should be resolved in a future.

\section{Summary}
In the framework of the q-1d picture of the FS proposed by us, 
we have calculated ${\rm Im} \chi(\vq,\,\omega)$ in the \lq RPA' 
within the slave-boson mean-field approximation to the $t$--$J$ model. 
We have found that ${\rm Im} \chi(\vq,\,\omega)$ shows 
two kinds of sharp peaks, the IC-peak  and the DIC-peak,  in both 
the u-RVB state and the $d$-RVB state. Their positions 
do not change appreciably with $\omega$  
and the sharp peaks survive down to low $\delta$. We have 
shown that the $d$-wave gap suppresses both the  
IC-peak and the DIC-peak, and that the former sees the magnetic gap
$\omega_{\rm mg}$ while the latter does not; 
interestingly  the latter is
more suppressed than the former for $\omega \sgt \omega_{\rm mg}$. 
We have also performed calculations for the 2dFS, and have found that 
the IC,DIC-peaks are realized 
only in the $d$-RVB and are replaced by the essentially C-peak in the u-RVB. 
This feature is crucially different from the results for the q-1dFS. 
Taking these results, 
we have argued that essential features of 
magnetic excitation in LSCO systems can be
understood in terms of the q-1d picture of the FS. 
Our scenario is different from the \lq spin-charge stripes' scenario 
where it is the formation of \lq charge stripes', not effects of the FS, 
that gives rise to the magnetic IC,DIC-peaks.

\section*{Acknowledgments} 
We thank H. Fukuyama for enlightening discussions and 
his continual encouragement. 
We also thanks H. Kimura, H. Kino, N. Ichikawa, M. Matsuda, 
M. Murakami, M. Saito, S. Uchida and  K. Yamada 
for helpful discussions.  
This work is supported by a Grant-in-Aid for Scientific Research on 
Priority Areas from Ministry of Education, Culture, Sports, Science 
and Technology, Japan. 

\newpage

\appendix
\section{Formalism in Presence of Interlayer Hopping}
 We take a unit cell in which two CuO$_{2}$ planes, 
$A$-plane and $B$-plane in \fig{q1dpicture}, are included, and 
separate the Bravais lattice into $A$-sublattice and $B$-sublattice. 
The $t$--$J$ model with the interlayer hopping integral, $t_{\perp}$, is 
then given by 
\bea
H =& & -  \sum_{i,\,j,\, \sigma}^{\qquad A} t\,^{(l)}
 f_{i\,\sigma}^{A \dagger}b_{i}^{A}b_{j}^{A \dagger}f_{j\,\sigma}^{A} + 
   J \sum_{<i,j>}^{\qquad A}  \vS_{i}^{A} \cdot \vS_{j}^{A} \nonumber\\
  & & -  \sum_{i,\,j,\, \sigma}^{\qquad B} t\,^{(l)} 
 f_{i\,\sigma}^{B \dagger}b_{i}^{B}b_{j}^{B \dagger}f_{j\,\sigma}^{B} + 
   J \sum_{<i,j>}^{\qquad B}  \vS_{i}^{B} \cdot \vS_{j}^{B} \nonumber \\ 
   & & -  \sum_{i\in A,\,j\in B} t_{\perp}
 \left(f_{i\,\sigma}^{A \dagger}b_{i}^{A}b_{j}^{B
 \dagger}f_{j\,\sigma}^{B}+ {\rm h.c.}\right),  \label{tztJ} \\
 & &\sum_{\sigma}f_{i\,\sigma}^{A \dagger}f_{i\,\sigma}^{A}
          +b_{i}^{A \dagger} b_{i}^{A}=1, \hspace{2mm} \sum_{\sigma}f_{i\,\sigma}^{B \dagger}f_{i\,\sigma}^{B}
          +b_{i}^{B \dagger}b_{i}^{B}=1 \nonumber \\   
& &  {\rm \hspace{40mm} at\  each\  site\ of\ {\makebox A\! \mbox{-}}\ and\ 
{\makebox B \mbox{-}}sublattices}. 
     \label{tzconstraint}
\eea  
We neglect the interlayer magnetic coupling $J_{\perp}$ whose order 
is estimated as  $\sim 10^{-5}J$\cite{thio,peters,lyons}. 
Following the same procedure in \S 2.1, but adding the mean fields, 
$\langle \sum_{\sigma}f_{i\,\sigma}^{A \dagger}
f_{j \,\sigma}^{B}\rangle$ and  
$\langle b_{i}^{A \dagger}b_{j}^{B}\rangle$,  
we obtain the mean-field Hamiltonian for the fermion part: 
\be
H_{\rm MF}=\sum_{\vk}\left(f_{\vk\,\uparrow}^{A \dagger}\;\;
f_{-\vk\,\downarrow}^{A}\;\; f_{\vk\,\uparrow}^{B \dagger}\;\;
f_{-\vk\,\downarrow}^{B}\right)
\left(\begin{array}{cccc} 
\xi_{\vk}^{A} & -\Delta_{\vk}& \epsilon_{\vk}& 0 \\
-\Delta_{\vk} & -\xi_{\vk}^{A} & 0  & -\epsilon_{\vk} \\
\epsilon_{\vk}& 0 & \xi_{\vk}^{B} & -\Delta_{\vk}\\
0 & -\epsilon_{\vk}&  -\Delta_{\vk}& -\xi_{\vk}^{B} 
\end{array}\right)
\left(\begin{array}{cccc}
f_{\vk\,\uparrow}^{A}\\
f_{-\vk\,\downarrow}^{A \dagger} \\ 
f_{\vk\,\uparrow}^{B}\\
f_{-\vk\,\downarrow}^{B \dagger}
\end{array} \right)\, , \label{tzMFH}
\ee
where
\bea
\xi_{\vk}^{A}=&&F_{x}\left(\cos k_{x}+\alpha \cos k_{y}\right)+  
F^{'} \cos k_{x} \cos k_{y}- \mu \; ,\\
\xi_{\vk}^{B}=&&F_{x}\left(\alpha \cos k_{x}+ \cos k_{y}\right)+  
F^{'} \cos k_{x} \cos k_{y}- \mu \; ,\\
\Delta_{\vk}=&&-\frac{3}{4} J \Delta_{0}(\cos k_{x}-\cos k_{y})\; ,\\
\epsilon_{\vk}=&&-8t_{\perp}\left<b_{i}^{A \dagger}b_{j}^{B}\right>
\cos \frac{k_{x}}{2}\cos \frac{k_{y}}{2}\cos \frac{k_{z}}{2}\; .
\eea
(Note that $\vk$ is a 3-dimensional vector.)
The form factor of $\epsilon_{\vk}$ comes from the fact that 
$A$- and $B$-sublattices are relatively displaced by 
$[\frac{1}{2},\frac{1}{2},\frac{1}{2}]$ (tetragonal notation).
We approximate 
$\langle b_{i}^{A \dagger}b_{j}^{B}\rangle \approx \delta$ 
and take the values of $F_{x}$, $F^{'}$,$\Delta_{0}$, 
$\mu$ and $\alpha$ as the 
same ones determined in \S 2.1 for the band parameters, 
$t^{(1)}/J =4$, $t^{(2)}/t^{(1)}=-1/6$ and $t^{(3)}/t^{(1)}=0$. 
The value of $t_{\perp}$ is taken to be $0.05t^{(1)}$ so that 
band width of $\epsilon_{\vk}$ is about 0.1 times that 
of $\xi_{\vk}^{A}$ (or $\xi_{\vk}^{B}$)\cite{pickett}.

Using the Hamiltonian \eq{tzMFH}, 
we obtain the irreducible dynamical magnetic susceptibility 
$\chi_{0}(\vq,\,\omega)$: 
\bea
&&\chi_{0}(\vq,\,\omega) \no \\ 
=&& \frac{1}{16NN_{z}}\sum_{\vk}\left[C^{1 +}_{\vk,\,\vk+\vq}
 \left(\tanh \frac{\beta \lambda_{\vk}^{+}}{2} 
          -\tanh \frac{\beta \lambda_{\vk+\vq}^{+}}{2}\right)
    \frac{1}
 {\lambda_{\vk}^{+}+ \omega +{\rm i}\Gamma -\lambda_{\vk+\vq}^{+} }
\right.\no \\
&&+\frac{1}{2}C^{1 -}_{\vk,\,\vk+\vq}
 \left(\tanh \frac{\beta \lambda_{\vk}^{+}}{2} 
          +\tanh \frac{\beta \lambda_{\vk+\vq}^{+}}{2}\right)
  \left(\frac{1}
 {\lambda_{\vk}^{+}+ \omega +{\rm i} \Gamma +\lambda_{\vk+\vq}^{+}}
 +\frac{1}
  {\lambda_{\vk}^{+}- \omega -{\rm i} \Gamma +\lambda_{\vk+\vq}^{+}}\right)\no \\
&&+C^{2 +}_{\vk,\,\vk+\vq}
 \left(\tanh \frac{\beta \lambda_{\vk}^{-}}{2} 
          -\tanh \frac{\beta \lambda_{\vk+\vq}^{-}}{2}\right)
    \frac{1}
 {\lambda_{\vk}^{-}+ \omega +{\rm i}\Gamma  -\lambda_{\vk+\vq}^{-}} \no \\
&&+\frac{1}{2}C^{2 -}_{\vk,\,\vk+\vq}
 \left(\tanh \frac{\beta \lambda_{\vk}^{-}}{2} 
          +\tanh \frac{\beta \lambda_{\vk+\vq}^{-}}{2}\right)
  \left(\frac{1}
 {\lambda_{\vk}^{-}+ \omega + {\rm i} \Gamma +\lambda_{\vk+\vq}^{-}}
 +\frac{1}
  {\lambda_{\vk}^{-}- \omega - {\rm i} \Gamma 
+\lambda_{\vk+\vq}^{-}}\right)\no \\
&&+ C^{3 +}_{\vk,\,\vk+\vq}
 \left(\tanh \frac{\beta \lambda_{\vk}^{+}}{2} 
          -\tanh \frac{\beta \lambda_{\vk+\vq}^{-}}{2}\right)
  \left(\frac{1}
 {\lambda_{\vk}^{+}+ \omega +{\rm i} \Gamma -\lambda_{\vk+\vq}^{-}}
 +\frac{1}
  {\lambda_{\vk}^{+}- \omega -{\rm i} \Gamma 
-\lambda_{\vk+\vq}^{-}}\right)\no \\
&&+\left. C^{3 -}_{\vk,\,\vk+\vq}
 \left(\tanh \frac{\beta \lambda_{\vk}^{+}}{2} 
          +\tanh \frac{\beta \lambda_{\vk+\vq}^{-}}{2}\right)
  \left(\frac{1}
 {\lambda_{\vk}^{+}+ \omega +{\rm i} \Gamma +\lambda_{\vk+\vq}^{-}}
 +\frac{1}
  {\lambda_{\vk}^{+}- \omega - {\rm i} \Gamma +\lambda_{\vk+\vq}^{-}}\right)
     \right]\, , \label{3dxqwo}
\eea 
where $2N_{z}$ ($N$) is the total number of CuO$_{2}$ planes 
(lattice sites in each CuO$_{2}$ plane) and 
$\vk$-summation is taken in the region, 
$-\pi \leq k_{x},k_{y},k_{z} \leq \pi$, and 
\bea
&& \lambda_{\vk}^{\pm}=\sqrt{\left(
\frac{\xi_{\vk}^{A}+\xi_{\vk}^{B}\pm \sqrt{D_{\vk}}}{2}\right)^{2}
 +\Delta_{\vk}^{2}} \; , \label{tzE}\\
&& D_{\vk}=\left(
\xi_{\vk}^{A}-\xi_{\vk}^{B}\right)^{2}+4\epsilon_{\vk}^{2}\; ,
\eea
and 
\bea
& C_{\vk,\,\vk+\vq}^{1 \pm}=\frac{1}{2} & \left( 
1+ \frac{\left(\xi_{\vk}^{A}-\xi_{\vk}^{B}\right)
         \left(\xi_{\vk+\vq}^{A}-\xi_{\vk+\vq}^{B}\right)
           +4 \epsilon_{\vk}\epsilon_{\vk+\vq}}
  {\sqrt{D_{\vk}D_{\vk+\vq}}}\right) \no \\
  && \times\left(
1 \pm \frac{\left(\xi_{\vk}^{A}+\xi_{\vk}^{B}+\sqrt{D_{\vk}}\right)
         \left(\xi_{\vk+\vq}^{A}+\xi_{\vk+\vq}^{B}+\sqrt{D_{\vk+\vq}}\right)
           +4 \Delta_{\vk}\Delta_{\vk+\vq}}
  {4\lambda_{\vk}^{+}\lambda_{\vk+\vq}^{+}}\right)\; , \\
& C_{\vk,\,\vk+\vq}^{2 \pm}=\frac{1}{2}& \left(
1+\frac{\left(\xi_{\vk}^{A}-\xi_{\vk}^{B}\right)
         \left(\xi_{\vk+\vq}^{A}-\xi_{\vk+\vq}^{B}\right)
           +4 \epsilon_{\vk}\epsilon_{\vk+\vq}}
  {\sqrt{D_{\vk}D_{\vk+\vq}}}\right) \no \\ 
 && \times \left(
1 \pm \frac{\left(\xi_{\vk}^{A}+\xi_{\vk}^{B}-\sqrt{D_{\vk}}\right)
         \left(\xi_{\vk+\vq}^{A}+\xi_{\vk+\vq}^{B}-\sqrt{D_{\vk+\vq}}\right)
           +4 \Delta_{\vk}\Delta_{\vk+\vq}}
  {4\lambda_{\vk}^{-}\lambda_{\vk+\vq}^{-}}\right)\; , \\
& C_{\vk,\,\vk+\vq}^{3 \pm}=\frac{1}{2} &\left(
1-\frac{\left(\xi_{\vk}^{A}-\xi_{\vk}^{B}\right)
         \left(\xi_{\vk+\vq}^{A}-\xi_{\vk+\vq}^{B}\right)
           +4 \epsilon_{\vk}\epsilon_{\vk+\vq}}
  {\sqrt{D_{\vk}D_{\vk+\vq}}}\right) \no \\
 && \times \left(
1 \pm \frac{\left(\xi_{\vk}^{A}+\xi_{\vk}^{B}+\sqrt{D_{\vk}}\right)
         \left(\xi_{\vk+\vq}^{A}+\xi_{\vk+\vq}^{B}-\sqrt{D_{\vk+\vq}}\right)
           +4 \Delta_{\vk}\Delta_{\vk+\vq}}
  {4\lambda_{\vk}^{+}\lambda_{\vk+\vq}^{-}}\right)\; .   
\eea
The \lq RPA' dynamical magnetic susceptibility is still given by  
\eq{RPA} and we set $r=0.35$. 
The positive infinitesimal value of $\Gamma$ is replaced with 
$0.01J$ as discussed in \S2.2.

In the numerical calculation of \eq{3dxqwo},   
we 
keep $2N_{z}=24$ CuO$_{2}$ planes to save computing time. 
The momentum $k_{z}$ is then discrete with a 
interval $2\pi/N_{z}$. 
From the sequence of the calculations with $N_{z}=1,4,8,12,25$, 
we expect that the overall $\vq$-dependence of ${\rm Im} \chi(\vq,\,\omega)$,
including the double-peak structures shown in \fig{tz}(d), 
and the locations of the IC,DIC-peaks do not depend on $N_{z}$ for 
$N_{z} \ge 8$.

\section{Analytic Formulae for Incommensurability and Magnetic Gap} 
We give formulae to estimate the peak positions of
${\rm Im} \chi(\vq,\,\omega)$  for a single CuO$_{2}$ plane, 
namely the values of $\eta_{x}$ and $\eta_{xy}$, at low 
$\omega$ and $T$.  From \eq{wmg}, 
the magnitude of the magnetic gap 
at $\eta_{y}$-peak is then calculated with such formulae. 

One of the $d$-wave gap nodes on the FS is estimated to be 
$\vk_{\rm  node}=-\frac{1}{2}(\pi-2\pi\eta_{xy},\,\pi-2\pi\eta_{xy})$, 
where 
\be
\sin(\pi \eta_{xy})=
\frac{F_{x}(1+\alpha)+\sqrt{F_{x}^2(1+\alpha)^{2}+4 \mu F^{'}}}
{-2 F^{'}}\; .\label{etaxy}
\ee
The $\eta_{y}$-peak position with the scattering vector 
$\vq=(\pi,\,\pi - 2 \pi \eta_{y})$ will be calculated by 
the minimum position of 
$\omega_{\rm mg}(\vq)=E_{\vq+\vk_{\rm node}}$,   
namely $\frac{\partial \omega_{\rm mg}(\vq)}{\partial \eta_{y}}=0$. 
We then obtain 
\be
\sin (2\pi\eta_{y}-\pi\eta_{xy})=\frac{(\alpha F_{x}-F^{'}\sin
(\pi\eta_{xy}))~(F_{x}\sin
(\pi\eta_{xy})+\mu)-\frac{9}{16}\Delta_{0}^{2}\sin (\pi\eta_{xy})}
{(\alpha F_{x}-F^{'}\sin (\pi
\eta_{xy}))^{2}+\frac{9}{16}\Delta_{0}^{2}}\, . \label{etay}
\ee
Numerical error of $\eta_{y}$ ($\eta_{xy}$) is found to be 
less than $\sim 15$\% ($\sim 1$\%) for $\D \sgt 0.04$ and almost 
vanishes for $\D \sgt 0.10$. Since we have checked that 
eqs.~(\ref{etaxy}) and (\ref{etay})   
reproduce the peak positions of ${\rm Im} \chi_{0}(\vq,\,\omega)$ 
quite well, the larger error at lower $\D$ is understood as coming 
from the $\vq$-dependence of the \lq RPA' enhancement factor 
(denominator in \eq{RPA}),  whose 
effects become prominent near the instability toward the
antiferromagnetic long-range order.  

The magnetic gap at $\eta_{y}$-peak  
is then calculated from eqs.~(\ref{wmg}), (\ref{etaxy}) and
(\ref{etay}),  
and plotted as a function of $\D$ in \fig{omegamg}.  It is 
seen that the analytical estimation reproduces the correct 
values (filled circles in \fig{omegamg}) for $\D \sgt 0.04$. 
The larger error at the lower $\D$ is due to the numerical error 
of $\eta_{y}$.

Here we note that \eq{etay} will be reduced to 
\be
\sin(2\pi\eta_{y}-\pi\eta_{xy})=\frac{ F_{x}\sin(\pi \eta_{xy})+ \mu }
{\alpha F_{x}-F^{'}\sin(\pi\eta_{xy})}\; , \label{etay2}
\ee
at high $\D$, since the band width $(\sim 2|F_{x}|)$ becomes  much 
larger than the $d$-wave singlet order $\Delta_{0}$. 
In fact, we find that the numerical error of $\eta_{y}$ estimated 
by \eq{etay2} is less 
than $\sim 5$\% for $\D \sgt 0.10$ and decreases to zero at the higher $\D$. 

When we employ eqs.~(\ref{etaxy}) and (\ref{etay2}), 
\eq{wmg} is reduced to 
\be
 \omega_{\rm mg}(\vq)=\frac{3}{4}J \Delta_{0}\left(\sin(\pi\eta_{xy})
-\sin(\pi \eta_{xy}-2\pi\eta_{y})\right)\; .  \label{wmg2} 
\ee
This simple formula 
is reasonable 
at least for high $\D$ $(\sgt 0.10)$.


\begin{thebibliography}{99}
\bibitem{Dthesis}  More precisely, the superlattice peaks 
are observed only at $(4\pi \pm 4\pi \D,\,0)$, and the extinction 
at $(4\pi,\, \pm 4\pi \D)$ has been argued  
as due to structure factors\protect\cite{tranquada2}.
\bibitem{xD} We consider $\D$ to be equal to the Sr$^{2+}$ concentration, 
	$x$.
\bibitem{magneticgap} The magnetic gap is often referred to the \lq spin
	gap'. In this paper, however, 
	we use the \lq spin gap' as a terminology
	indicating the normal state gap.
\end{thebibliography}
\end{document}